\documentclass[11pt]{article}
\usepackage{amsmath,amsfonts,amssymb,eucal}
\advance \textheight by \topskip 
\makeatletter
\@addtoreset{equation}{section}
 \def\theequation{\thesection.\arabic{equation}}
\makeatother

\newtheorem{theorem}{Theorem}[section]

\newtheorem{definition}[theorem]{Definition}

\newtheorem{proposition}[theorem]{Proposition}
\newtheorem{remark}[theorem]{Remark}

\def\ZZ{\mathbb Z}
\def\CC{\mathbb C}
\def\RR{\mathbb R}
\def\KK{\mathbb K}
\def\HH{\mathbb H}
\def\FF{\mathbb F}
\def\C{{\cal  C}}
\def\tvi{\vrule height 12pt depth 6pt width 0pt}
\def\tv{\tvi\vrule}
\def\cc#1{\kern .7em\hfill #1 \hfill\kern .7em}
\newcommand{\beq}{\begin{equation}}
\newcommand{\eeq}{\end{equation}}
\newcommand{\beqa}{\begin{eqnarray}}
\newcommand{\eeqa}{\end{eqnarray}}
\newcommand{\e}{\varepsilon}

\newcommand{\noi}{\noindent}

\def\>{\rangle}
\def\<{\langle}

\date{\today}

\begin{document}
\begin{titlepage}

\title{
{\bf Clifford Algebras in  Physics}}

\author{
{\sf M. Rausch de Traubenberg}\thanks{e-mail:
rausch@lpt1.u-strasbg.fr}\,\,
\\
{\small {\it
Laboratoire de Physique Th\'eorique, CNRS UMR  7085,
Universit\'e Louis Pasteur}}\\
{\small {\it  3 rue de l'Universit\'e, 67084 Strasbourg Cedex, France}}\\ 
}
\date{}
\maketitle
\vskip-1.5cm

\vspace{2truecm}

\begin{abstract}
\noindent
We study briefly some properties of real Clifford algebras and
identify them as matrix algebras.
We then show that the representation space on which  Clifford algebras act are spinors
and we  study in details   matrix representations.
The precise structure of these matrices gives rise to the type of spinors
one is able to construct in a given space-time dimension: Majorana or Weyl.
Properties of spinors are also studied.
We finally show how Clifford algebras enable us to construct supersymmetric extensions
of the Poincar\'e algebra. A  special attention to the four, ten and eleven-dimensional
space-times is given. We then study  the representations of the considered supersymmetric
algebras and show that representation spaces contain an equal number of bosons and
 fermions. Supersymmetry turns out to be a symmetry which mixes non-trivially the
bosons and the fermions since one multiplet contains bosons and fermions together.
We also show how supersymmetry in four and ten dimensions  are related to eleven dimensional
supersymmetry by compactification or dimensional reduction.
\end{abstract}

\bigskip \bigskip \bigskip

\noi
{\it Lecture given at the Preparatory Lectures of the $7^{\text{th}}$ International
Conference on Clifford Algebras and their Applications $-$ICCA7$-$, Toulouse, France,  May 19-29, 2005.}
\end{titlepage}
\newpage
\begin{titlepage}
\tableofcontents
\end{titlepage}
\newpage

\vspace{2cm}

\newpage
\section{Introduction}
\renewcommand{\theequation}{1.\arabic{equation}}   
\setcounter{equation}{0}
The real  Clifford algebra  $\C_{t,s}$ is the associative algebra generated
by a unit $1$ and  $d=t+s$ elements $e_1,\cdots,e_d$ satisfying

\beqa
e_M e_N + e_N e_M = 2 \eta_{MN}, \ \ \text{with }
\eta_{MN}= \left\{ 
\begin{array}{ll}
\ \;  0& \text{if } M\ne N \\
\ \;1& \text{if }M=N=1,\cdots, t \\
-1&\text{if } M=N=t+1,\cdots,d.
\end{array} \right.
\eeqa

\noi
It is well-known that the  Clifford  algebra $\C_{t,s}$ can be represented by  
$2^{\left[\frac{d}{2}\right]} \times 2^{\left[\frac{d}{2}\right]} $
matrices called the
 Dirac matrices (with $[a]$ the integer part of $a$). Since $\mathfrak{so}(t,s) \subset \C_{t,s}$,
the  $2^{\left[\frac{d}{2}\right]}-$dimensional
complex vector space on which the Dirac matrices act is also a representation of   $\mathfrak{so}(t,s)$, the 
spinor representation. Thus on a physical ground Clifford algebras are intimately related to
fermions. The four dimensional Clifford algebra, or more precisely its matrix representation,
was introduced in physics by Dirac \cite{dirac} 
 when he was looking for a relativistic first order differential
equation extending the Schr\"odinger equation.  This new equation called now, the Dirac equation,
is in fact a relativistic equation describing spin one-half fermions as the electron. 

Among the various applications of Clifford algebras in physics and
mathematical physics, we will show how
they are  central tools  in the construction of supersymmetric theories. 
In particle physics there are  two types of particles:   bosons and  fermions.
The former, as the photon,  have integer spin and the latter, as the electron, have
half-integer spin. Properties (mass, spin, electric charge {\it etc.}) or the way particles
interact together are understood by means of Lie algebras (describing space-time 
and internal symmetries). Supersymmetry is a symmetry different from the previous ones
in the sense that it is a symmetry which mixes bosons and fermions. Supersymmetry
is not described  by  Lie algebras but  by Lie superalgebras and contains generators in the spinor representation
of the space-time symmetry group. 

In this lecture, we will show, how 
supersymmetry is intimately related to Clifford algebras. Along all the steps of the construction, a large
number of details will be given. Section 2 will be mostly mathematical and devoted to the definition,
and the classification of  Clifford algebras. In section 3, we will show how  Clifford
algebras are related to special relativity and to spinors. Section 4 is a technical section, central
for the construction of supersymmetry. Some basic properties of the Dirac $\Gamma-$matrices (the
matrices representing  Clifford algebras) will be studied allowing to introduce different types
of spinors  (Majorana, Weyl). We will show how the existence of these types of spinors is
crucially related to the
space-time dimension. Useful details will be summarized in some tables.  Finally, in section 5 we will
construct explicitly supersymmetric algebras and their representations with a special emphasis 
to the four, ten and eleven-dimensional space-time. We will also show how lower dimensional
supersymmetry are related to higher dimensional supersymmetry by dimensional reduction.  
\section{Definition and classification}
\renewcommand{\theequation}{2.\arabic{equation}}   
\setcounter{equation}{0}
In this section we  give the definition of Clifford algebras. We then show that it is possible
to characterize Clifford algebras as matrix algebras and show that the properties of 
real Clifford algebras depend on the dimension modulo $8$.

\subsection{Definition}
Consider $E$ a $d-$dimensional vector space over the field $\mathbb K$
($\mathbb K = \RR$ or $\mathbb K= \CC$). Let $Q$ be a quadratic form, 
non-degenerate of signature $(t,s)$ on
$E$. The quadratic form $Q$ naturally defines a symmetric scalar product
on $E$

\beqa
\label{prod-scal}
\eta(x,y)=\frac14(Q(x+y)-Q(x-y)), \mbox{~for all~}x,y \mbox{~in~} E.
\eeqa

\noi
Denote  $\eta=\text{Diag}(\underbrace{1,\cdots,1}_t,\underbrace{-1,\cdots,-1}_s)$
the tensor metric in an orthonormal basis. Denote also $\eta_{MN}, \ 1 \le M,N \le d $ the 
matrix elements of $\eta$ and $\eta^{MN}$ the matrix elements of the inverse matrix 
$\eta_{MN}=\eta^{MN}= \left\{\begin{array}{ll} 0, & M\ne N \cr
                                               1, &M=N=1,\cdots,t \\
                                               -1, &M=N=t+1,\cdots.d\end{array} \right.$ .


\begin{definition}\label{def-cliff}
1) The associative $\KK-$algebra  ${\cal C}^\KK_{t,s}$    
generated by a unit  $1$ and 
$d=t+s$ generators $e_M, M=1,\cdots,t+s$  satisfying 

\beqa
\label{clif}
\left\{e_M,e_N\right\}=e_M e_N + e_N e_M = 2 \eta_{MN} 
\eeqa

\noi
is called the Clifford algebra of the quadratic form $Q$.

2) The dimension of ${\cal C}^\KK_{t,s}$ is $2^d$ and a convenient basis
is given by

\beqa
\label{basis}
1, e_{M_1}, e_{M_1} e_{M_2}, \cdots, e_{M_1} e_{M_2} \cdots e_{M_{d-1}},
e_1 e_2 \cdots e_d, \
1 \le M_1 < M_2 < \cdots < M_{d-1} \le d.
\eeqa
\end{definition}

\begin{remark}
The case where $Q(x)=0$ for all $x$ in $\KK$
will be considered latter on (see remark \ref{weyl}) and corresponds to Grassmann algebras.
\end{remark}

\begin{remark}
\label{clifford-formal}
Clifford algebras could have been defined in a more formal way as
follow. Let $T(E)= \KK \oplus E \oplus \left( E\otimes E\right) \oplus
\cdots$ be the tensor algebra over $E$ and let ${\cal I}(Q)$ be the two-sided ideal
generated by $x\otimes x -Q(x)1, x \in E$ in $T(E)$. The quotient algebra
$T(E)/{\cal I}(Q)$ is the Clifford algebra ${\cal C}^\KK_{t,s}$.
The canonical map $\imath_Q : E \to {\cal C}^\KK_{t,s}$ given by the
composition $E\to T(E) \to  {\cal C}^\KK_{t,s}$ is an injection.
\end{remark}

\begin{remark}
\label{exterior}
If we set $\Lambda(E)= \KK \oplus  E  \oplus \Lambda^2(E) \oplus \cdots \oplus \Lambda^d(E)$ 
to be the exterior algebra on $E$, 
 then the canonical map $\imath_Q$ extends to a
vector-space isomorphism $\hat \imath_Q : \Lambda(E) \to \C_{t,s}^\KK$ given on $\Lambda^n(E)$ by

$$\hat \imath_Q(v_1 \wedge \cdots \wedge v_n) =
\sum\limits_{\sigma\in S_n} \frac {1}{n!}
 v_{\sigma(1)}\dots v_{\sigma(n)}$$

\noi
with $S_n$ the group of permutations with $n$ elements.
\end{remark}
\subsection{Structure}
We call $\C_0{}^\KK_{t,s}$ 
the subalgebra  of  $\C^\KK_{t,s}$ 
generated by an even  product of generators $e_M$ and 
$\C_1{}^\KK_{t,s}$ the vector space (which is not a subalgebra)  of  $\C^\KK_{t,s}$ 
generated by an  odd product of generators $e_M$.
We also introduce

\beqa
\label{epsilon}
\varepsilon = e_1 e_2 \cdots e_d
\eeqa

\noi
corresponding to the product of all the generators of $\C_{t,s}^\KK.$
An easy calculation gives 

\beqa
\label{epsilon2}
\e^2 = (-1)^{\frac{d(d-1)}{2}} e_1^2 \cdots e_M^2 = (-1)^{\frac{d(d-1)}{2}+s} =
\left\{ \begin{array}{ll} 
(-1)^{\frac{t-s}{2}}&d \text{\ even} \cr
(-1)^{\frac{t-s-1}{2}}& d \mbox{\ odd.}
\end{array} \right.
\eeqa

\noi
Furthermore, when $d$ is  even $\e$ commutes with $\C_0{}^\KK_{t,s}$ 
and anticommutes with $\C_1{}^\KK_{t,s}$ and when $d$ is  odd  $\e$ 
commutes both with $C_0{}^\KK_{t,s}$ and
$\C_1{}^\KK_{t,s}$. It can be shown that  the centre $\ZZ(\C_{t,s}) = \KK$ 
(resp. $\ZZ(\C_{t,s} ) = \KK \oplus \KK \, \e$)  when $d$ is even (resp. when $d$ is odd).
The element $\e$ will be useful to decide whether or not  Clifford algebras are simple.
 We first characterize   real Clifford algebras
that we denote from now on  ${\cal C}_{t,s}$ to simplify  notation.
Then, we study the case of  complex Clifford algebras
${\cal C}_{t,s}^{ \CC} = {\cal C}_{t,s} \otimes_\RR \CC$.

\subsubsection{Real Clifford algebras}
From \eqref{epsilon2} $\e^2=1$ for $t-s= 0 \ (\text{mod. } 4)$ when
$d$ is even and $\e^2=1$ for $t-s=1 \ (\text{mod. }4)$ when $d$ is odd and thus 
in these cases $P_\pm =\frac12 (1 \pm \e)$ 
define orthogonal projection operators $(P_+ P_-=0, P_+ + P_1=1)$.  These projectors will enable us to define ideals of
$\C_{t,s}$ and $\C_0{}_{t,s}$ \cite{ABS}.

\begin{proposition} 
\label{simple-real}
Let $\C_{t,s}$ be a real Clifford algebra and let $P_\pm$ be defined as above.
\begin{itemize}
\item[(i)]
When $d$ is  odd  and $t-s=1 \ (\text{mod. }4)$ then $P_\pm$ belong to the centre 
$\ZZ(\C_{t,s})$ and   $\C_{t,s}= P_+ \C_{t,s} \oplus  P_- \C_{t,s}$. This means that
$\C_{t,s}$  is not simple
($P_\pm  \C_{t,s}$ are two ideals of $C_{t,s})$. 
\item[(ii)] When $d$ is odd, $\C_0{}_{t,s}$ is  simple.
\item[(iii)]
When $d$ is even, $\C_{t,s}$ is simple.
\item[(iv)]
When $d$ is even  and $t-s=0 \text{ (mod. }4)$ then $P_\pm$ belong to the centre 
$\ZZ(\C_0{}_{t,s})$ and thus we have $\C_0{}_{t,s}= P_+ \C_0{}_{t,s} \oplus  P_- \C_0{}_{t,s}$. 
This means that  $\C_0{}_{t,s}$  is not simple 
($P_\pm  \C_0{}_{t,s}$ are two ideals of $C_0{}_{t,s})$.
\end{itemize}
\end{proposition}
The proof will in fact be given in table \ref{tab1}, section 4.

\subsubsection{Complex Clifford algebras}
In this case, $E$ is a complex vector space, and the tensor metric $\eta$ can always be chosen to be
Euclidien, {\it i.e.} $\eta = \mathrm{Diag}(1,\cdots,1)$. We denote now
$\overline{\C}_d$ a  complex Clifford algebras generated by $d$ generators. When we are over the field of complex
numbers,  a projection operator can always be defined. Indeed if $\e^2=-1, (i\e)^2=1$,
so we set $P_\pm = \frac12(1 \pm \e)$ if $\e^2=1$ and 
$P_\pm = \frac12(1 \pm i \e)$ if $\e^2=-1$.

\begin{proposition}
\label{simple-complexe}
Let $\overline{\C}_d$ be a complex Clifford algebra, we have the following structure:
\begin{itemize}
\item[(i)] when $d$ is odd, $\overline{\C}_d= P_+ \overline{\C}_d \oplus  P_-\overline{ \C}_d$ 
is not simple;
\item[(ii)] when $d$ is even, $\overline{\C}_d$ 
is  simple;
\item[(iii)] when $d$ is odd, $\overline{\C}_0{}_d$  is  simple;
\item[(iv)] when $d$ is even, $\overline{\C}_0{}_d= P_+ \overline{\C}_0{}_d \oplus  
P_-\overline{ \C}_0{}_d$  is not simple.
\end{itemize}
\end{proposition}

\noi

\subsection{Classification}
\label{classification}
We now show that 
it is possible to characterize all Clifford algebras as matrix algebras
\cite{ABS, coq}. (See also \cite{coq2} and references therein.) 
 We first study the case of real Clifford algebras.
The interesting point in this classification   is
two-fold. Firstly, the knowledge of 
${\cal C}_{1,0}, {\cal C}_{0,1},{\cal C}_{2,0},{\cal C}_{0,2},{\cal C}_{1,1}$
determines all the other ${\cal C}_{t,s}$. Secondly, the properties of
${\cal C}_{t,s}$ depend on $t-s$ mod. $8$.

\subsubsection{Real Clifford algebras}
In these family of algebras we may add ${\cal C}_{0,0} = \RR$.
From the definition \ref{def-cliff} one can show 

\beqa
\label{small-cliff}
\begin{array}{lll}
{\cal C}_{0,0} = \RR & {\cal C}_{1,0} = \RR \oplus \RR &{\cal C}_{0,1} = \CC \cr
{\cal C}_{2,0} = {\cal M}_2(\RR) &{\cal C}_{1,1} =  {\cal M}_2(\RR) &{\cal C}_{0,2}= \HH
\end{array}
\eeqa 

\noi
with $\HH$ the quaternion algebra and 
${\cal M}_n(\FF)$  the $n \times n$ matrix algebra over the field $\FF= \RR, \CC$ or $\HH$.
Recall that the quaternion algebra is generated by three imaginary units $i,j,k=ij$ satisfying
$i^2=j^2=-1$ and $ij +ji =0$ which is precisely  the definition of ${\cal C}_{0,2}$.
For the algebra $\C_{2,0}$ if we set $h_{11}=\frac12(1+e_1), h_{22}=\frac12(1-e_1), h_{12}=h_{11}e_2$
and $h_{21}=h_{22}e_2$ one can check that 
they are independent and that 
they satisfy the multiplication law of
${\cal M}_2(\RR)$.
Similar simple arguments give the identifications (\ref{small-cliff}).
Next, notice the

\begin{proposition}
We have the following isomorphisms
\label{prop-iso-cliff}
\beqa
\label{cliff-iso}
(1) \ \ {\cal C}_{t,s} \otimes_\RR {\cal C}_{2,0} \cong {\cal C}_{s+2,t}, \ \ 
(2) \ \ {\cal C}_{t,s} \otimes_\RR {\cal C}_{1,1}\cong {\cal C}_{t+1,s+1}, \ \ 
(3) \ \ {\cal C}_{t,s} \otimes_\RR {\cal C}_{0,2} \cong {\cal C}_{s,t+2}.
\eeqa
\end{proposition}
\noi

\noi
{\bf Proof:}
To prove the first isomorphism, introduce $\left\{e_M\right\}_{M=1,\cdots,d}$ 
(resp. $\left\{f_1,f_2\right\}$) the generators of ${\cal C}_{t,s}$  (resp. ${\cal C}_{2,0}$).
Then, one can show that $\left\{g_M=e_M \otimes f_1 f_2, g_{d+1}= 1\otimes f_1,g_{d+2}= 1\otimes f_2 \right\}_{M=1,\cdots,d}$
(with $1$ the unit element of ${\cal C}_{t,s}$) satisfy the relations \eqref{clif} for
${\cal C}_{s+2,t}$.  To end the proof we just have to check that 
the $g_{M_1} \cdots g_{M_\ell}, 1\le M_1 < \cdots < M_\ell \le d+2,
\ \ell =1, \cdots, d+2$ are independent.
A similar proof works for the two other isomorphisms. { {\it QED}}
 
\medskip
Finally, we recall some well-known isomorphisms of real  matrix algebras

\begin{proposition}
\beqa
\label{matrix-iso}
\begin{array}{ll}
(1) \ \ {\cal M}_m(\RR) \otimes{\cal M}_n(\RR) \cong {\cal M}_{mn}(\RR),&
(5) \ \ \CC \otimes_\RR \CC  \cong \CC \oplus \CC,\cr
(2) \ \ {\cal M}_m(\RR) \otimes \RR \cong {\cal M}_m(\RR),&
(6) \ \  \CC \otimes_\RR \HH \cong {\cal M}_2(\CC), \cr
(3) \ \ {\cal M}_m(\RR) \otimes_\RR \CC \cong {\cal M}_m(\CC),&
(7)\ \ \HH \otimes_\RR \HH \cong {\cal M}_4(\RR).  \cr
 (4) \ \ {\cal M}_m(\RR) \otimes_\RR \HH \cong {\cal M}_m(\HH),
\end{array} 
\eeqa
\end{proposition}

\noi
{\bf Proof:}  The only non-trivial isomorphisms to be proved  are those involving  the quaternions.
Recall that
  $\HH \cong \Big\{ q_0 \sigma_0 + q_1 (-i \sigma_1) + q_2 (-i\sigma_2) +
q_3 (-i \sigma_3), q_0,q_1,q_2, q_3 \in \RR \Big\}$ with 

\beqa
\label{pauli}
\sigma_0 = \begin{pmatrix}1&0\cr 0&1\end{pmatrix}, \ \ 
\sigma_1 = \begin{pmatrix}0&1\cr 1&0\end{pmatrix}, \ \ 
\sigma_2 = \begin{pmatrix}0&-i\cr i&0\end{pmatrix}, \ \ 
\sigma_3 = \begin{pmatrix}1&0\cr 0&-1\end{pmatrix},
\eeqa

\noindent
the Pauli matrices, is a faithful representation of the quaternions in ${\cal M}_2(\CC)$. Thus, $(6)$ is easily proven.

To prove (7), let $q= q_0 1 + q_1 i + q_2 j + q_3 k, 
q'= q'_0 1 + q'_1 i + q'_2 j + q'_3 k $ be two given quaternions and set
$\bar q' =   q'_0 1 - q'_1 i - q'_2 j - q'_3 k$. Define now

\beqa
f_{q,q'} : \HH &\rightarrow &\HH \nonumber \\
      x&  \mapsto &x'=qx\bar q'. \nonumber 
\eeqa 

\noi
It is now a matter of calculation to show that $f_{q,q'}$ can be represented by a $4 \times 4$
real matrix acting on the component of the quaternion $x$. This proves $(7)$. { {\it QED}}

From the isomorphisms (\ref{cliff-iso}) and (\ref{matrix-iso}) one is able to calculate
all the algebras ${\cal C}_{d,0}$ and ${\cal C}_{0,d}$. As an illustration, we give the result
for $C_{0,7}:$
  
\beqa
\C_{0,7} &\cong &\C_{5,0} \otimes \C_{0,2} \cong \C_{0,3} \otimes \C_{2,0} \otimes \C_{0,2} 
\cong  \C_{1,0}\otimes \C_{0,2} \otimes \C_{2,0} \otimes \C_{0,2}  \nonumber \\
&\cong& \left(\RR \oplus \RR\right) \otimes \HH \otimes {\cal M}_2(\RR) \otimes \HH \nonumber \\
&\cong&\left(\HH \oplus \HH\right) \otimes {\cal M}_2(\RR) \otimes \HH \nonumber \\
&\cong&\left({\cal M}_4(\RR) \oplus {\cal M}_4(\RR) \right) \otimes  {\cal M}_2(\RR) \nonumber \\
&\cong&{\cal M}_8(\RR) \oplus {\cal M}_8(\RR). \nonumber
\eeqa

\noi
The Clifford algebras $\C_{d,0}, \C_{0,d}, 0\le d \le 8$ are given in table \ref{tab1}.

\begin{table}[!ht]
$$\vbox{\offinterlineskip \halign{
\tv# & \cc{#} & \tv# & \cc{#}  & \tv# &
\cc{#} & \tv# & \cc{#} & \tv#  \cr
\noalign{\hrule}
&\cc{$d$}&&$\C_{d,0}$ &&$\C_{0,d}$& \cr
\noalign{\hrule}
&\cc{$1$}&&$\RR \oplus \RR$ &&$\CC$& \cr
\noalign{\hrule}
&\cc{$2$}&&${\cal M}_2(\RR)$ &&$\HH$& \cr
\noalign{\hrule}
&\cc{$3$}&&${\cal M}_2(\CC)$ &&$\HH \oplus \HH$& \cr 
\noalign{\hrule}
&\cc{$4$}&&${\cal M}_2(\HH)$ &&${\cal M}_2(\HH)$& \cr  
\noalign{\hrule}
&\cc{$5$}&&${\cal M}_2(\HH) \oplus {\cal M}_2(\HH)$ && ${\cal M}_4(\CC)$&  \cr 
\noalign{\hrule}
&\cc{$6$}&&${\cal M}_4(\HH)$ &&${\cal M}_8(\RR)$& \cr
\noalign{\hrule}
&\cc{$7$}&&${\cal M}_8(\CC)$ &&${\cal M}_8(\RR) \oplus {\cal M}_8(\RR)$& \cr
\noalign{\hrule}
&\cc{$8$}&&${\cal M}_{16}(\RR)$ &&${\cal M}_{16}(\RR)$& \cr
\noalign{\hrule}
}}$$
\caption{{\small \underline{Clifford algebras $\C_{d,0}$ and $\C_{0,d}$, $1\le d \le 8$.}}}\label{tab1}
\end{table}

\noindent
We observe as claimed in the previous subsection that only $\C_{1,0}, \C_{0,3}, \C_{5,0}, \C_{0,7}$
are not simple  (see Proposition \ref{simple-real} (i)).

Next the identity

\beqa
\label{mod8}
\C_{0,2} \otimes \C_{2,0}\otimes \C_{0,2}\otimes \C_{2,0} \cong \C_{0,8} \cong \C_{8,0} \cong
{\cal M}_{16}(\RR)
\eeqa

\noi
yields 
\beqa
\C_{d+8,0} \cong \C_{d,0} \otimes \C_{8,0}, \ \ \C_{0,d+8} \cong \C_{0,d} \otimes \C_{0,8}.
\eeqa

\noi 
Thus if $\C_{d,0} \cong {\cal M}_n(\FF)$, ($\FF=\RR, \CC$ or $\HH$), then
$\C_{d+8,0} \cong {\cal M}_{16n}(\FF)$.
This ends the classification of the algebras ${\cal C}_{d,0}$ and ${\cal C}_{0,d}$
 \cite{ABS}.  The case of space-time with arbitrary signature \cite{coq} are obtained 
from $\C_{d,0}$ and $\C_{0,d}$ through the identity

\beqa
\C_{t,s} &\cong& \C_{t-s,0} \otimes \underbrace{\C_{1,1}\otimes \cdots  \otimes \C_{1,1}}, \ \ t-s \ge 0\nonumber  \\
&& \hskip 2.truecm s \mbox{~times}  \\
&\cong& \C_{t-s,0} \otimes {\cal M}_{2^s}(\RR), \ \ \ t >s  \nonumber 
\eeqa

\noi
since $\C_{1,1} \cong {\cal M}_2(\RR).$
Thus if $\C_{t-s,0} \cong {\cal M}_n(\FF)$, ($\FF=\RR, \CC$ or $\HH$), then
$\C_{t,s} \cong {\cal M}_{2^sn}(\FF)$.
 We summarize in table \ref{tab2} the results for the
Clifford algebras $\C_{t,s}$ with an arbitrary signature.

\begin{table}[!ht]
$$\vbox{\offinterlineskip \halign{
\tv# & \cc{#} & \tv# & \cc{#}  & \tv# &
\cc{#} & \tv#  \cr
\noalign{\hrule}
&\cc{$t-s$ mod. $8$}&&$\C_{t,s}$ &\cr
\noalign{\hrule} 
&\cc{$0$}&&${\cal M}_{2^\ell}(\RR)$ & \cr
\noalign{\hrule}
&\cc{$1$}&&${\cal M}_{2^\ell}(\RR)\oplus {\cal M}_{2^\ell}(\RR)$ & \cr
\noalign{\hrule}
&\cc{$2$}&&$ {\cal M}_{2^\ell}(\RR)$& \cr 
\noalign{\hrule}
&\cc{$3$}&&$ {\cal M}_{2^\ell}(\CC)$& \cr  
\noalign{\hrule}
&\cc{$4$}&&$ {\cal M}_{2^{\ell-1}}(\HH)$ & \cr 
\noalign{\hrule}
&\cc{$5$}&&$ {\cal M}_{2^{\ell-1}}(\HH)\oplus {\cal M}_{2^{\ell-1}}(\HH)$& \cr
\noalign{\hrule}
&\cc{$6$}&&$ {\cal M}_{2^{\ell-1}}(\HH)$& \cr
\noalign{\hrule}
&\cc{$7$}&&${\cal M}_{2^{\ell}}(\CC) $&\cr
\noalign{\hrule}
}}$$
\caption{{\small \underline{ Clifford algebras $\C_{t,s}$.}\ \ \  ($\ell = [d/2]$, the integer part of
$d/2, \ d=t+s$).}}\label{tab2}
\end{table}

\noi
The only non-simple algebras are those
when $t-s=1 \text{ mod. }4$.

\subsubsection{Complex Clifford algebras}
Finally we give the classification for  complex Clifford algebras. The {\it complex} Clifford
algebras $\C_{t,s}^\CC \cong \C_{t,s} \otimes_\RR \CC$
are obtained by  complexification of the {\it real} Clifford algebras $\C_{t,s}$. Of course
the complexification of the algebra $\C_{t,s}$ depends only on $d=t+s$ and
it will be denoted $\overline{\C}_d$. Using tables \ref{tab1} and \ref{tab2} we 
obtain

\beqa
\label{complex-cliff}
\overline{\C}_d=\left\{ 
\begin{array}{ll}
{\cal M}_{2^{\left[\frac{d}{2}\right]}}(\CC)& d \mbox{~even} \cr
{\cal M}_{2^{\left[\frac{d}{2}\right]}}(\CC)\oplus {\cal M}_{2^{\left[\frac{d}{2}\right]}}(\CC)& 
d \mbox{~odd}.
 
\end{array} \right.
\eeqa


\section{Clifford algebras in relation to special relativity}
\label{relat}
After recalling the definition of the basic group of invariance of special relativity, we show that 
the representation
spaces on which  Clifford algebras act  correspond to spinors. We then obtain in a natural way the
group $\text {Pin}(t,s)$ which is a non-trivial double covering of the group $O(t,s)$. We further identify
the Lie algebra $\mathfrak{so}(t,d) \subset \C_{t,s}$.
\renewcommand{\theequation}{3.\arabic{equation}}   
\setcounter{equation}{0}
\subsection{Poincar\'e and Lorentz groups}
The basic group of special relativity ({\it i.e} in a four dimensional  Minkowskian space-time with
a metric of signature $(1,3)$) can be easily extended to a $d-$dimensional
pseudo-Euclidien space of signature $(t,s)$. 
Let $i_i, \cdots, i_{t+s}$ be an orthonormal basis of the vector space $E$. The
directions $i_1, \cdots, i_t$ are time-like  ($\eta(i_M,i_M)=1>0,\  1 \le M \le t$)
and the directions $i_{t+1},\cdots,i_{s+t}$ are space-like ($\eta(i_M,i_M)=-1< 0,\  t+1 \le M \le t+s$) .
The Lorentz group  
is defined by 
$O(t,s)=\left\{ f \in GL(E) : 
\eta(f(x), f(y))= \eta(x,y), \forall x,y \in E \right\}$. 
If $\Lambda$ denotes a matrix representation of $f$,
from $f(x)=\Lambda x, \eta(x,y)=x^t \eta y$ and $\eta(f(x),f(y))=\eta(x,y)$, we get 
$\Lambda^t \eta \Lambda = \eta$
\noi with $\Lambda^t$ the transpose of the matrix $\Lambda$, and $\Lambda$ is
a  transformation  preserving the tensor metric.
The Lorentz group has four connected components

\beqa
\label{lorentz2}
{\cal L}=O(t,s)= {\cal L}_+^{\uparrow} \oplus {\cal L}_+^{\downarrow}
\oplus {\cal L}_-^{\uparrow} \oplus {\cal L}_-^{\downarrow}
\eeqa

\noi
where ${\cal L}_\pm^{\uparrow},{\cal L}_\pm^{\downarrow} $  represent elements  of determinant
$\pm 1$ and $\uparrow$ (resp. $\downarrow$) represents elements with positive (negative) temporal
signature. 
Let $R_i, \ i=1,\cdots, d$ be the reflections in the hyperplane perpendicular to
the $i^{\text{th}}$ direction. $R_1,\dots,R_t$ are time-like reflections and
$R_{t+1},\cdots R_{t+s}$ are space-like reflections. More generally one can consider
a reflection $R(v)$ orthogonal to  a given direction $v \in E$ such that $\eta(v,v) \ne 0$ 
said to be time-like if $\eta(v,v)>0$
and space-like if $\eta(v,v)<0$. 
An element of $O(t,s)$ is given by a products of certain numbers of such reflections. 
The structure of the various components of the Lorentz group are as follow

\beqa
\label{lorentz3}
\begin{array}{ll}
{\cal L}_+^\uparrow,  &\text{ is a continuous group, {\it i.e.} is  associated to some 
Lie algebra } \cr
{\cal L}_-^\uparrow = R(v) {\cal L}_+^\uparrow,& \text{ where } v \text{ is a space-like direction,
say } i_{t+1} \cr
 {\cal L}_-^\downarrow = R(v) {\cal L}_+^\uparrow, &\text{ where  } v \text{ is a time-like direction,
say } i_{1}  \cr
{\cal L}_+^\downarrow = R(v) R(v') {\cal L}_+^\uparrow, &\text{ where  } v \text{ is a time-like direction,
say } i_{1}  \cr
& \text{ and  where  } v' \text{ is a space-like direction,
say } i_{t+1}. 
\end{array}
\eeqa

\noi
In other words, if $ R(v_1) \cdots R(v_n)$  is a product  of 
(i)  an even number of space-like and time-like reflections 
it belongs to ${\cal L}_+^\uparrow$, 
(ii) an odd number of space-like and an  even number of time-like reflections 
it belongs to  ${\cal L}_-^\uparrow$, 
(iii) an even number of space-like and an odd number of time-like reflections 
it belongs to  ${\cal L}_-^\downarrow$, 
(iv)  an odd number of space-like and time-like reflections to
it belongs to ${\cal L}_+^\downarrow$.
When the signature is $(1,d-1)$,  $R(e_1)=T$ is the operator of
time reversal (it changes the direction of the time), 
and when in addition $d$ is even  $R(e_2)  \cdots R(e_d)=P$ is the parity operator
(it corresponds to a reflection with respect to the origin).
Furthermore, one can identify several subgroups of $O(t,s)$: 

\beqa
\label{sub-lorentz}
O(t,s) = {\cal L}, \ \ \ SO(t,s)= {\cal L}_+^\uparrow \oplus {\cal L}_+^\downarrow, \ \ \
SO_+(t,s)= {\cal L}_+^\uparrow.
\eeqa

\noi
$SO(t,s)$ is constituted of an even product of reflections and $SO_+(t,s)$ an even product
of time-like and space-like reflections.
Note that with an Euclidien metric, $O(d)$ has only two connected components. 
Moreover, none of these groups  are simple connected. 

For further use, we introduce the generators
of $\mathfrak{so}(t,s)$ the Lie algebra of $SO_+(t,s)$. A conventional basic
is given by the $L_{MN}=-L_{NM}$ which corresponds to the generators 
which generate the  ``rotations'' in the plane $(M-N)$.
We also introduce  $P_M$ ($1 \le M \le d$)
the generators of the space-time translations.\footnote{If $x^M,  \ M=1, \cdots,d$ denote the  
components in the $d-$dimensional vector space $E$ and we set $x_N = \eta_{NM}x ^M$,  
 we have $L_{MN}=( x_M \frac{\partial \ \ \  }{\partial x^N} -  x_N \frac{\partial \ \ \  }{\partial x^M})$ and 
$P_M =  \frac{\partial  \ \ \ }{\partial x^M}$.}   
The generators $L_{MN}$ and $P_M$  generate the so-called Poincar\'e
algebra or the inhomogeneous Lorentz algebra  noted $\mathfrak{iso}(t,s)$ and satisfy 

\beqa
\label{poincare}
\left[L_{MN}, L_{PQ}\right]&=&
-\eta_{MP} L_{NQ}+\eta_{NP} L_{MQ} - \eta_{MQ}L_{PN}+\eta_{NQ} L_{PM}, \nonumber \\
\left[L_{MN}, P_P \right]&=& \eta_{NP} P_M -\eta_{MP} P_N.
\eeqa

\subsection{Universal covering group of the Lorentz group}
In this section we only consider  real Clifford algebras $\C_{t,s}$. We
furthermore identify a vector $v = x^M i_M\in E$ with an element
$x^M e_M \in \C_{t,s}$ (see remark \ref{clifford-formal}). With such an identification,
we have $x^2=((x^1)^2 + \cdots + (x^t)^2 -(x^{t+1})^2 - \cdots- (x^{t+s})^2= \eta(x,x).$

\begin{definition} The Clifford group $ \Gamma_{t,s}$ 
is the subset of invertible elements $x$ of $\C_{t,s}$ such that 
$\forall v \in E, xvx^{-1} \in E$.
\end{definition}

\noi
It is clear that invertible elements of $E$ belong to $ \Gamma_{t,s}$
(this excludes the null vectors {\it i.e } the vectors such that $\eta(v,v)=0$).
 If  we take $x \in E \subset \C_{t,s}$ invertible, since $xv+vx=2 \eta(x,v)$, we have 

\beqa
x v x^{-1} = \frac{1}{x^2}xvx =-\left(v-\frac{2 \eta(x,v)}{x^2} x \right)= -R(x)(v),
\eeqa

\noi
which corresponds to a symmetry in the hyperplane perpendicular to $x$.
More generally, for $s \in \Gamma_{t,s}$, the transformation $\rho(s)$ defined by
$\rho(s)(x)=sxs^{-1}$ belong to $O(t,s)$.
However, the representation $\rho:  \Gamma_{t,s} \to GL(E)$ is not faithful because if $x \in \Gamma_{t,s}$,
then $ a x \in \Gamma_{t,s}$  $ (a \in \CC^\star)$ and $\rho(ax)=\rho(x)$.
A standard way to distinguish the elements $x$ and $ax$ is to introduce a normalisation. For that purpose
we define\ \  $\bar{  \text{  } }: \C_{t,s} \to \C_{t,s}$ by

\beqa
\label{beta}
 \overline{e_M}=(-1)^{t+1} e_M, \  \ \
\overline{e_{M_1}. \cdots .e_{M_{k-1}}. e_{M_k}}=\overline{e_{M_k}} \ . \ 
\overline{e_{M_{k-1}}} \ . \ 
\cdots \ . \ \overline{e_{M_1}}, 
\eeqa

\noi
  (see remark \ref{adjoint}). Next, we define 
 $N(x)= x \bar x$. Note that
in the case of the quaternions ($t=0,s=2$), we have $\bar i=-i, \bar j=-j, \bar k=-k$.
Now, with this definition of the norm, for $x \in E$ we have 

\beqa
\label{norm}
(-1)^{t+1} N(x)= x^2 \left\{\begin{array}{ll} 
 >0 &\text{ if }  x \text{ is time-like } \cr 
 < 0&\text{ if }  x \text{ is space-like } 
\end{array}\right.
\eeqa

\begin{proposition} (Proposition 3.8 of \cite{ABS})
If $x \in \Gamma_{t,s}$ then $N(x) \in \mathbb R^\star$.
\end{proposition}

\noi 
As a consequence for $x,y\in \Gamma_{t,s}$ we have $N(xy)=N(x) N(y)$.
Then, the ``norm'' $N$ enables us to define definite subgroups of the Clifford group $\Gamma_{t,s}$.
The first subgroup which can be defined is 

\beqa
\text{Pin}(t,s)&=& \Big\{x \in \Gamma_{ t,s } \text{ s.t. } \vert N(x) \vert =1\Big\}.
\eeqa

\noi
By construction it is easy to see that $\rho(\Gamma_{t,s})= \rho(\text{Pin}(t,s))$ 
and that $x \in E$ invertible $\Rightarrow x \in \text{Pin}(t,s)$.
Moreover, since an element of $O(t,s)$  is given by a product of a given number of reflections,
we have $\rho(\text{Pin}(t,s)) \supseteq O(t,s)$. 
Conversely it has been shown (\cite{ABS}, proposition 3.10) that
$\rho(\text{Pin}(t,s)) \subseteq O(t,s)$,  
thus $\rho(\text{Pin}(t,s)) = O(t,s)$. A generic element of $\text{Pin}(t,s)$ is then given by

\beqa
\label{Pin}
s= v_1 v_2 \cdots v_n
\eeqa

\noi
with $v_i, i=1,\cdots,n$ invertible elements of $E$
(thus $\rho(s)$ corresponds to the transformation given by
 $R(v_1)R( v_2) \cdots R(v_n)$). Assume now, that there are $p$ time-like vectors
and $q$ space-like vectors  in $s$ ($p+q=n$), then $N(s)=(-1)^{p(t+1)+qt}=(-1)^{nt+p}.$

\begin{enumerate}
\item If $n$ is  even  then $\rho(s) \in SO(t,s)$.
\item If $n$ is even  and $N(s) >0$ then $\rho(s) \in SO_+(t,s)$.
\item If $n$ is even and $N(s) <0$ then $\rho(s) \in {\cal L}_+^\downarrow$.
For instance when $d$ is an even number and $t=1, s=d-1$,
$\rho(\e)= \rho(e_1 \cdots e_d) \in {\cal L}_+^\downarrow$
and corresponds to the $PT$ inversion ($P$ being the operator of parity transformation
 and $T$ the operator of time reversal.) 
\item If $n$ is  odd  and $N(s)(-1)^{t+1} >0$ then $\rho(s)  \in {\cal L}_-^\downarrow$.
For instance when $(t,s)=(1,d-1), \rho(e_1) \in {\cal L}_-^\downarrow$ and corresponds
to the operator of time reversal.
\item If $n$ is  odd  and $N(s)(-1)^{t+1} <0$ then $\rho(s)  \in {\cal L}_-^\uparrow$.
For instance when $d$ is even and $(t,s)=(1,d-1), \rho(e_1 \cdots e_d)) \in {\cal L}_-^\uparrow$ and 
corresponds to the operator of parity transformation.
\end{enumerate}

\noi
We can now define the various subgroup of $\Gamma_{t,s}$:
\beqa
\text{Pin}(t,s)&=& \Big\{x \in \Gamma_{ t,s } \text{ s.t. } \vert N(x) \vert =1\Big\} \nonumber \\
\text{Spin}(t,s)&=& \text{Pin}(t,s) \cap \C_0{}_{ t,s }  \\ 
\text{Spin}_+(t,s)&=& \Big\{x \in \text{Spin}(t,s) \text{ s.t. }  N(x) =1\Big\} \nonumber
\eeqa  


\subsection{The Lie algebra of $\text{Spin}_+(t,s)$}
Among the various groups of the previous subsection, only $\text{Spin}_+(t,s)$ is a connected Lie
group. If one introduce the $d(d-1)/2$ elements

\beqa
\label{spin}
S_{MN}= \frac14(e_M e_N -e_N e_M), \ \ 1\le M \ne N \le d
\eeqa

\noi
a direct calculation gives

\beqa
\label{spin2}
\left[S_{MN}, e_P\right]&=& \eta_{NP} e_M - \eta_{MP} e_N \\
\left[S_{MN}, S_{PQ} \right]&=&-\eta_{MP} S_{NQ}+\eta_{NP} S_{MQ} - \eta_{MQ}S_{PN}+\eta_{NQ} S_{PM}.
\nonumber
\eeqa

\noi
This means that $S_{MN}$ generate the Lie algebra $\mathfrak{so}(t,s)$ ($\mathfrak{so}(t,s) \subset \C_{t,s}$) and that
$e_M$ are in the vector representation of $\mathfrak{so}(t,s)$. 
Furthermore, if we define

\beqa
\label{p-form1}
e^{(\ell)}_{M_1 \cdots M_\ell}= \frac{1}{\ell !} \sum \limits_{ \sigma \Sigma_\ell}  \epsilon(\sigma)
e_{M_{\sigma(1)}} \cdots e_{M_{\sigma(\ell)}},
\eeqa

\noi
(with $\epsilon(\sigma)$ the signature of the permutation $\sigma$)
 using
\eqref{spin2} one can show that they are in the $\ell^{\text{ th}}-$antisymmetric representation
of $\mathfrak{so}(t,s)$ (see remark \ref{exterior}).

\begin{proposition}
 The group $\text{Spin}_+(t,s)$ is a non-trivial
double covering group of $SO_+(t,s).$
\end{proposition} 

\noi 
{\bf Proof:} 
First notice that if $x \in \text{Spin}_+(t,s)$ then $-x \in  \text{Spin}_+(t,s)$. Then, we 
show that there exist a continuous path in $ \text{Spin}_+(t,s)$ which connects $1$ to $-1$.
Let $i_M,i_N$ be two space-like directions.
The path  $R(\theta) =e^{\frac12 \theta e_M e_N}= \cos \frac{\theta}{2} + \sin \frac{\theta}{2} e_M e_N$
with $ \theta \in [0, 2\pi]$ is such that $R(0)=1$ and $R(2 \pi)=-1$ and thus connects $1$ to $-1$ in
$ \text{Spin}_+(t,s)$.
Which ends the  proof. {\it QED}

In the same way, $\text{Spin}(t,s)$ is a non-trivial  double covering group of $SO(t,s)$ and
$\text{ Pin}(t,s)$\footnote{This is a joke of J.- P. Serre.} a non-trivial  double covering group of $O(t,s)$.

\section{The Dirac $\Gamma-$matrices}  
\label{spinors}
\renewcommand{\theequation}{4.\arabic{equation}}   
\setcounter{equation}{0}
In this section, because we are mostly interested in  Clifford 
algebras in relation
to space-time physics, we will focus on vector spaces with a Lorentzian signature $(1,d-1)$ or
an Euclidian signature $(0,d)$. From now on, in the case of the $(1,d-1)$ 
signature, we
use the notations commonly used in the literature. The indices of the 
space-time
components run from $0$ to $d-1$, greek indices $\mu,\nu, \cdots=0,\cdots,d-1$
 are space-time indices
and latin indices $i,j,\cdots=1,\cdots,d-1$ are space indices.
This section is devoted to the study of the  matrix representation of  
Clifford algebras.
The precise structure of these matrices gives rise to the type of spinors
one is able to construct in a given space-time dimension: Majorana or Weyl.
Properties of spinors are also studied.
This section is technical but is central
for the construction of supersymmetric theories in various dimensions.

\subsection{Dirac spinors}
For  physical applications, we need to have matrix representations of  Clifford algebras.
As we have seen in section \ref{classification}, the case where $d$ is 
 even is
very different from the case where $d$ is odd. Indeed,  complex
 Clifford algebras
are simple when $d$ is even and are not simple when $d$ is odd. This means 
that the representation
is faithful when $d$ is even  and not faithful  when $d$ is odd. When $d$ is 
odd,
since  $\e = e_1 \cdots e_d$ allows to define the two ideals of the Clifford 
algebra, $\e$ will
be represented by a number. Moreover, one of the ideals will be the kernel of 
the representation. 

It is well-known that the basic building block of the matrix representation of $\bar C_{d}$
are the Pauli matrices \eqref{pauli} \cite{BW}.  Define the ($2k+1$) matrices
by the tensor products of $k$ Pauli matrices (we thus construct $2^k \times 2^k$ matrices):

\beqa
\label{gamma}
&&\begin{array}{ll}
\Sigma_1^{(k)} = \sigma_1 \otimes  \underbrace{\sigma_0 \otimes \cdots \otimes \sigma_0}_{k-1},
& 
\Sigma_2^{(k)} = \sigma_2 \otimes \underbrace{\sigma_0 \otimes \cdots \otimes \sigma_0}_{k-1}, \cr
\Sigma_3^{(k)} = \sigma_3 \otimes \sigma_1 \otimes
 \underbrace{\sigma_0 \otimes \cdots \otimes \sigma_0}_{k-2},& 
\Sigma_4^{(k)} = \sigma_3 \otimes \sigma_2 \otimes  
\underbrace{\sigma_0 \otimes \cdots \otimes \sigma_0}_{k-2}, \cr
\hskip1.6truecm\vdots & \hskip1.5truecm \vdots\cr
\Sigma_{2 \ell -1}^{(k)}  = \underbrace{\sigma_3 \otimes \cdots \otimes \sigma_3}_{\ell-1}
\otimes \sigma_1 \otimes  \underbrace{\sigma_0 \otimes \cdots \otimes \sigma_0}_{k-\ell},
&  
\Sigma_{2 \ell}^{(k)} = \underbrace{\sigma_3 \otimes \cdots \otimes \sigma_3}_{\ell-1}
 \otimes \sigma_2 \otimes 
 \underbrace{\sigma_0 \otimes \cdots \otimes \sigma_0}_{k-\ell},\cr
\hskip1.6truecm  \vdots & \hskip1.5truecm  \vdots\cr
\Sigma_{2 k -1}^{(k)}  = \underbrace{\sigma_3 \otimes \cdots \otimes \sigma_3}_{k-1}
\otimes \sigma_1,&
\Sigma_{2 k}^{(k)} = \underbrace{\sigma_3 \otimes \cdots \otimes \sigma_3}_{k-1},
\otimes \sigma_2, \cr
\end{array} \nonumber \\
&& \hskip 4.5truecm
\Sigma_{2k+1}^{(k)}= \underbrace{\sigma_3 \otimes \cdots \otimes \sigma_3}_{k}.
\eeqa

\noi
Observing that

\beqa
\label{rec} 
\Sigma^{(k+1)}_M=\Sigma^{(k)}_M \otimes\sigma_0 ,\  M=1,\cdots,2k,\ 
\Sigma^{(k+1)}_{2k+i}=\Sigma_{2k+1}^{(k)}\otimes \ \sigma_i,\ i=1,2,3 
\eeqa

\noi
a simple recurrence 
on $k$
shows that the $\Sigma-$matrices 
satisfy $\Sigma_M^{(k)} \Sigma_N^{(k)} + \Sigma_N^{(k)} \Sigma_M^{k)} = 2 \delta_{MN}$. Thus

\beqa
\label{rep-mat}
\gamma :&\bar \C_{d}& \to {\cal M}_{2^k}(\CC) \nonumber  \\
&e_M& \mapsto \Sigma_M
\eeqa

\noi
($d=2k$ or $d=2k+1$)
is a representation of $\bar C_{d}$.  This representation is faithful 
when $d$ is even and  non-faithful  when
$d$ is odd ($\gamma(\e)=\Sigma_1^{(k)} \Sigma_2^{(k)} \cdots \Sigma_{2k+1}^{(k)} = i^{k}.$)
Since the group $\text{Spin}_+(t,s)$ is the double covering of the group $SO_+(t,s)$,
the  representation on which the $\Gamma-$matrices act is a  representation  of
$\text{Spin}_+(t,s)$. The elements of the representation space $\bar \C_{d}$ are called
Dirac spinors. A Dirac spinor exists in any dimension $d$ and has $2^{\left[\frac{d}{2}\right]}$
complex components. We denote by $\Psi_D$ a Dirac spinor.

Now having represented the algebra $\bar \C_d$, ($d=2k$ or $d=2k+1$) if we set 
$\Gamma_M= \Sigma_M^{(k)}, \ 1 \le M \le t$ 
and $\Gamma_M= i \Sigma_M^{(k)},\  t+1 \le M \le t+s$ we have  a representation of the algebra
$\C_{t,s}$ corresponding to 
a real form of $\bar \C_d$.
In particular  for a Minkowskian space-time we introduce 

\beqa
\label{Gamma(1,d-1)}
\Gamma_0 = \Sigma_1^{(k)}, 
\Gamma_j= i \Sigma_{j+1}^{(k)}, j=1,\cdots,d-1.
\eeqa

\noi
We also denote by

\beqa 
\label{spin-generators} 
\Gamma_{\mu \nu} =\frac14(\Gamma_\mu \Gamma_\nu -\Gamma_\nu \Gamma_\mu)
\eeqa

\noi  the
generators of the  Lie algebra $\mathfrak{so}(1,d-1)$.

\begin{remark}
\label{weyl}
Without using the result of the section \ref{classification} we show that 
$\bar \C_{2n} \cong {\cal M}_{2^n}(\CC)$ as follows. Set

\beqa
\label{fermi}
a_i=\frac12\left(e_{2i} + i e_{2i+1}\right), \ \ 
b_i= \frac12\left(e_{2i} - i e_{2i+1}\right), \ \ i=1,\cdots, n
\eeqa

\noi
which satisfy 

\beqa
\label{grassmann}
a_i a_j + a_j a_i =0, \ \
b_i b_j + b_j b_i =0, \ \
a_i b_j+ b_j a_i= \delta_{ij}.
\eeqa

\noi
Thus the $a_i$ and $b_i$ generate the fermionic  oscillator algebra ({\it i.e.} the algebra
which underlines the fermionic fields after quantization).
Note that the $a_i$ alone generate the Grassmann algebra of dimension $n$.
To obtain a representation of the algebra \eqref{grassmann}, we introduce the 
Clifford vacuum $\Omega= \left|-\frac12,-\frac12,\cdots,-\frac12\right>$ 
such that $a_i \Omega=0, i=1,\cdots,n$. Then we obtain the representation
of the algebra \eqref{grassmann} by acting in all possible
 ways with $b_i$ at most once each:

\beqa
\label{rep}
\left|s_1,s_2,\cdots,s_n\right>= (b_1)^{s_1+1/2} \cdots (b_n)^{s_n+1/2} 
\left|-\frac12,-\frac12,\cdots,-\frac12\right>, \ \ s_1,s_2,\cdots,
s_n=-\frac12, \frac12
 \eeqa

\noi
and thus we obtain a $2^n-$dimensional representation.
The notation with $s_i=-\frac12, \frac12$ 
(instead of $s_i=0,1$) seems to be unnatural, but is
appropriate to identify the weight of $\left|s_1,s_2,\cdots,s_n\right>$
with respect to  the Cartan subalgebra of $\mathfrak{so}(2n,\CC)$.
Furthermore, it can be shown that
$a_i^\dag = b_i$ and that the representation is unitary for the real algebra $\C_{2n,0}$.
In this case, the notation reflects the property of the Dirac spinor
 with respect to the group
$\text{Spin}(2n)$. A basis of the Cartan subalgebra of $\mathfrak{so}(2n)$
can be taken to be
$\Gamma_{2i\  2i+1}= i a_i a_i^\dag -\frac{i}{2}$ and the vector
$\left|s_1,s_2,\cdots,s_n\right>$ is an eigenvector of $\Gamma_{2i \  2i+1}$ with eigenvalue
$is_i$. The half-integer weights show that we have a spinor representation of $\text{Spin}(2n)$.
(The $i$ factor comes from the fact that $SO(2n)$ is the real form of $SO(2n,\CC)$ corresponding
to the maximal compact algebra, for which the generators are antihermitian $-$note that, in the physical
literature there is no $i$ factor since  
$\Gamma_{\mu \nu} = i \frac12 \Gamma_\mu \Gamma_\nu$ 
instead of \eqref{spin-generators}.$-$ ) To end this remark we just notice that the fermionic
generators $(a_i,b_i)$ are in the vector representation of $\text{Spin} (2N)$.

\end{remark}

\subsection{Majorana and Weyl spinors}
In the previous subsection we have introduced Dirac spinors. In this subsection,
we will see that further spinors can be defined such as Majorana spinors or Weyl spinors 
{\it etc}. The Weyl spinors are just a consequence of the reducibility of the (Dirac)
spinors representation of $\mathfrak{so}(2n,\CC)$ (and of course of all its real forms).
They exist in any even space-time dimensions and for any signature.
If there exists a real matrix representation of the Clifford algebra $\C_{t,s}$ then
one is able to consider real (or Majorana) spinors. The existence of Majorana spinors
depends on the space-time dimension and of the signature of the metric.

If we denote ${\cal S}$ the vector space corresponding to the Dirac spinors,
recall the Lorentz generators writes \eqref{spin}

\beqa
\label{lorentz-gamma}
\Gamma_{\mu \nu}= \frac14(\Gamma_\mu \Gamma_\nu- \Gamma_\nu \Gamma_\mu).
\eeqa

\noi
The matrices $\Gamma_\mu$ act on the representation ${\cal S}$ and the representation
of $\C_{t,s}$ is just $\text {End}({\cal S})$, the set of endomorphisms of ${\cal S}$.
In the same way the matrices $-\Gamma^t$ (with ${}^t$ the transpose
operation) act on the dual representation ${\cal S}^\star$,
the matrices $\Gamma^\star$ (with ${}^\star$ the complex conjugation\footnote{In the
mathematical literature the complex conjugate of $\Gamma_\mu$ is denoted $\bar \Gamma_\mu$.})
act on  the complex-conjugate representation $\bar {\cal S}$ and
the matrices $\Gamma^\dag$ (with ${}^\dag$ the hermitian conjugation)
act on the representation $\bar {\cal S}^\star$.
These representations are in fact equivalent, since we can find elements of $\text {End}({\cal S})$
such that

\beqa
\label{equiv}
A \Gamma_{\mu \nu} A^{-1}= -\Gamma_{\mu \nu}^\dag, \
B \Gamma_{\mu \nu} B^{-1}= \Gamma_{\mu \nu}^\star, \
C \Gamma_{\mu \nu} C^{-1}= -\Gamma_{\mu \nu}^t
\eeqa

\noi
(see below). The operators $A,B,C$ are intertwining operators
($B$ intertwines the representations ${\cal S}$ and $\bar {\cal S}$).
This is in fact related to \eqref{lorentz-gamma} and

\beqa
\label{equiv-gamma}
A \Gamma_{\mu} A^{-1}= \eta_A\Gamma_{\mu}^\dag, \
B \Gamma_{\mu } B^{-1}= \eta_B\Gamma_{\mu }^\star, \
C \Gamma_{\mu } C^{-1}= \eta_C\Gamma_{\mu \nu}^t,
\eeqa

\noi 
with $\eta_A,\eta_B$ and $\eta_C$  signs which depend on $d$, as we will see in  subsection \ref{majoranasect}.

As a direct consequence of \eqref{equiv},  if  $\Psi_D \in {\cal S}$ transforms like 
$\Psi_D'=S(\alpha) \Psi_D=e^{\frac12 \alpha^{\mu \nu} \Gamma_{\mu \nu}}
\Psi_D$ under a Lorentz transformation 
($\alpha_{\mu \nu} \in \RR$ are the parameters of the transformation)
then 
$\bar \Xi_D= \Xi_D^\dag A$ and  $\Xi^t C$ belong to ${\cal S}^\star$
({\it i.e} transform with $S(\alpha)^{-1}$). Thus, $\bar \Xi_D\Psi_D$ and
$\Xi_D^t C \Psi_D$ define invariants.







%

\subsubsection{Weyl spinors}
Consider first the case of the complex  Clifford algebra $\bar \C_d$.
When $d$ is  even  the spinor representation  ${\cal S}$ is reducible
${\cal S} = {\cal S}_+ \oplus {\cal S}_-$. Indeed, if  we define the chirality matrix 

\beqa
\label{chirality}
\chi=  i^{\left[\frac{d}{2}\right]+1}\Gamma_0 \Gamma_1 \cdots \Gamma_{d-1}
\eeqa

\noi
it satisfies 

\beqa
\label{chirality2}
\chi^2=1, \ \left\{\Gamma_\mu, \chi\right\}=0, \ 
\left[\chi, \Gamma_{\mu \nu}\right]=0.
\eeqa

\noi
Hence, $\chi$ allows to define the complex left- and right-handed Weyl spinors.
These spinors correspond to the two irreducible representations of $\bar \C_d$:

\beqa
\label{weyl-spinors}
\Psi_L = \frac12(1-\chi) \Psi_D, \ 
\Psi_R = \frac12(1+\chi) \Psi_D,
\eeqa 

\noi
$\Psi_L \in {\cal S}_-, \Psi_R \in {\cal S}_+$ and  $\Psi_D \in {\cal S}$ or 
${\cal S}_\pm = \left\{ \Psi \in {\cal S} \text { s.t. } \chi \Psi= \pm \Psi \right\}.$
This means that the spaces ${\cal S}_\pm$ carry an irreducible representation of the
complex algebra $\bar \C_{0d}$. Now, if we concider the real form $\C_0{}_{t,s}$ of
$\bar \C_{0d}$,  the spaces ${\cal S}_\pm$ become  irreducible
representations of $\C_0{}_{t,s}$. The generators of the two
Weyl spinors of $\mathfrak{so}(1,d-1)$ are $\Sigma^\pm_{\mu \nu}= \frac12 \left(1 \pm \chi\right) \Gamma_{\mu \nu}$.
The Weyl  spinors are called the semi-spinors
in the mathematical literature.

\begin{remark}
As we have seen (see Proposition \ref{simple-complexe} (i)), the algebra $\bar \C_0{}_d$
 is not simple. Then it falls into two simple ideals $\C_0{}_d =P_+\C_0{}_d+ P_- \C_0{}_d$.
Moreover, the Lie algebra $\mathfrak{so}(t,s) \subset \C_0{}_d$, this means that a
(complex) Dirac spinor is reducible into two (complex) Weyl spinors.
\end{remark}
\subsubsection{Majorana spinors}
\label{majoranasect}
Majorana spinors are real spinors.
As we now see, the existence of Majorana spinors crucially depends on the dimension $d$ and
on the metric signature. We  consider the case of Lorentzian signatures
$(1,d-1)$.  The general case can be easily deduced. However, 
for a general study see
\cite{gso}, \cite{kt}.

The key observation in this subsection is the simple fact that (i) the 
Pauli matrices are hermitian (ii) the matrices $\sigma_1,\sigma_3$
are real and symmetric and (iii)  the matrix $\sigma_2$ is purely imaginary and antisymmetric.
Thus from \eqref{gamma} we see (take $d=2n+1$ odd)

\beqa
\label{conj}
\begin{array}{llll}
\Gamma_0^\dag=\Gamma_0,&\Gamma_0^\star=\Gamma_0,&\Gamma_0^t=\Gamma_0,& \cr
\Gamma_{2i}^\dag=-\Gamma_{2i},&\Gamma_{2i}^\star=-\Gamma_{2i},&\Gamma_{2i}^t=\Gamma_{2i},& 
i=1,2,\cdots, n\cr
\Gamma_{2i-1}^\dag=-\Gamma_{2i-1},&\Gamma_{2i-1}^\star=\Gamma_{2i-1},&\Gamma_{2i-1}^t=-\Gamma_{2i-1},& 
i=0,2,\cdots,n.
\end{array}
\eeqa

\noi
The results that we will establish here  seems to depend
 of the choise of   basis we have chosen, but in fact  they
are independent of this choice (see \cite{gso}).

\medskip
  \framebox{ Space-time of even dimension $d=2n$}  \\

\noi  
When $d$ is even, we take the first $d$ matrices above. In this case, 
we have seen  that  complex Clifford algebras are isomorphic to
 some matrix algebras. This means that if we have another representation of the 
Clifford algebra $\Gamma_\mu' (\left\{\Gamma_\mu',\Gamma_\nu'\right\}= 2 \eta_{\mu \nu}$), there 
exists an invertible matrix $U$ of $\text{End} ({\cal S})$ such that $\Gamma_\mu'=U \Gamma_\mu U^{-1}$.
In particular,  we have (see \eqref{conj} and \eqref{equiv})

\beqa
\label{ABC}
\begin{array}{ll}
\chi \Gamma_\mu \chi^{-1}=-\Gamma_\mu,& \chi= i^{n+1} \Gamma_0 \Gamma_1 \cdots \Gamma_{2n-1}\cr
A \Gamma_\mu A^{-1}=\Gamma_\mu^\dag,& A=\Gamma_0 \cr
B_1 \Gamma_\mu B_1^{-1}=(-1)^{n+1 }\Gamma_\mu^\star,& B_1=\Gamma_2 \Gamma_4 \cdots
\Gamma_{2n-2}\cr
B_2 \Gamma_\mu B_2^{-1}=(-1)^{n}\Gamma_\mu^\star,& B_2=\Gamma_0 \Gamma_1 \cdots
\Gamma_{2n-1}\cr
C_1 \Gamma_\mu C_1^{-1}=(-1)^{n+1 }\Gamma_\mu^t,& C_1=\Gamma_0 \Gamma_2 \cdots
\Gamma_{2n-2} 
 \cr
C_2 \Gamma_\mu C_2^{-1}=(-1)^{n}\Gamma_\mu^t,& C_2=\Gamma_1 \Gamma_3 \cdots
\Gamma_{2n-1} \cr
\end{array}
\eeqa

\noi
$\Gamma_0$ is the only hermitian matrix, 
$B_1$ is the product of all purely imaginary matrices, $B_2$ is the product of
real matrices, $C_1$ is the product of symmetric matrices and $C_2$ of antisymmetric
matrices. Note also the relation between the matrices $A,B,C$: $C_i= A B_i, i=1,2$.
We set $\eta_1=(-1)^{n+1 }, \eta_2=(-1)^n$.
From the definition of $B$, we have

\beqa
\label{B}
B_1B_1^\star=(-1)^{\frac {(n-1)(n-2)}{2}}=\epsilon_1, \ \  
B_2B_2^\star =(-1)^{\frac {n(n-1)}{2}}=\epsilon_2. \ \ 
\eeqa

\noi
Now for further use, we collect the following signs

\beqa
\label{signs}
\begin{array}{lllll}
d=2 \text{ mod. }8 & \epsilon_1=+, &\eta_1=+& \epsilon_2=+, &\eta_2=- \cr
d=4 \text{ mod. }8  & \epsilon_1=+,& \eta_1=-& \epsilon_2=-,& \eta_2=+ \cr
d=6 \text{ mod. }8  & \epsilon_1=-,& \eta_1=+& \epsilon_2=-,& \eta_2=- \cr
d=8 \text{ mod. }8 &  \epsilon_1=-, &\eta_1=-& \epsilon_2=+, &\eta_2=+.
\end{array}
\eeqa
\noi

\medskip
\noi
\underline{Majorana spinors, $d$ even}
\medskip

\noi
From \eqref{ABC}, we get $B \Gamma_{\mu \nu} B^{-1} = \Gamma^\star_{\mu \nu},\  B=B_1,B_2,$
so the Dirac spinor $\Psi_D$ and $B^{-1} \Psi_{D}^\star$ transform in the same way under the group
$\text{Pin}(1,d-1)$. A Majorana spinor is a spinor that we impose to be  a {\it real} spinor. 
It satisfies 

\beqa
\label{majorana}
\Psi_M^\star=B \Psi_M, \ \ B=B_1, B_2.
\eeqa

\noi
But taking the complex conjugate of the above equation gives $\Psi_M = B^\star \Psi_M^\star=
B^\star B \Psi_M$. Thus this is possible only if $BB^\star=1$ or when $\epsilon_1=1$ and/or
$\epsilon_2=1$ and from \eqref{signs} when 
$d=2,4,8 \text{ mod. }8$. More precisely, looking to \eqref{ABC}, we observe that the $\Gamma-$matrices 
can be taken to be purely real if $\eta_1=\epsilon_1=1$ or $\eta_2=\epsilon_2=1$ {\it i.e.} 
if $d=2,8 \text{ mod. }8$ and the Dirac matrices can be taken purely imaginary if
$\epsilon_1=1, \eta_1=-1$ or $\epsilon_2=1, \eta_2=-1$ that is when $d=2, 4 \text { mod. }8$
(In this case  the matrices $\Gamma_\mu$ are purely imaginary, the matrices $\Gamma_{\mu \nu}^{(2)}$
are real, the matrices $\Gamma_{\mu \nu \rho}^{(3)}$ are purely imaginary {\it etc.}). 
The first type of
real spinors will be called  Majorana spinors although the second type of spinors  pseudo-Majorana
spinors. This result could have been deduced from table \ref{tab2}. Indeed, we have 
shown that $\C_{t,s}$ is real when $t-s=0,1,2 \text{  mod. }8 $. 
Observing that $\C_{t,s}$ is related to $\C_{s,t}$ by the transformation $e_M \to i e_M$, we get
that $\gamma(e_M)= \Gamma_M$ are  purely imaginary when $s-t=1,2 \text{ mod. } 8$. Thus we have: 

\begin{proposition}
\label{majorana-even}
Assume $d$ is even.
\begin{itemize}
\item[(i)]  Majorana spinors of ${\text Pin}(1,d-1)$ exist when $d=2,8 \text{ mod. }8$.
\item[(ii)] Pseudo-Majorana spinors of ${\text Pin}(1,d-1)$ exist when $d=2,4 \text{ mod. }8$.
\end{itemize}
\end{proposition}

\medskip
\noi
\underline{$SU(2)-$Majorana spinors, $d$ even}
\medskip

\noi
As we have seen, a Majorana spinor is a spinor which satisfies $\Psi_M=B \Psi_M^\star,$ with 
$BB^\star=1$.  However, if $BB^\star=-1$ one can define $SU(2)-$Majorana spinors (or
$SU(2)-$pseudo-Majorana spinors). This is a pair of spinors $\Psi_i, i=1,2$ satisfying

\beqa
\label{su2}
(\Psi_i)^\star= \epsilon^{ij} B \Psi_j,
\eeqa

\noi
where $\epsilon^{ij}$ is the $SU(2)-$invariant antisymmetric tensor, and $i,j=1,2$.
(More generally one can take an even number of spinors and substitute to $\epsilon$
the symplectic form $\Omega$. These spinors are also called symplectic spinors.)
The $SU(2)-$Majorana spinors exist when $\epsilon_1=-1, \eta_1=1$ or 
$\epsilon_2=-1, \eta_2=1$ and the the $SU(2)-$pseudo-Majorana spinors exist when 
$\epsilon_1=-1, \eta_1=-1$ or 
$\epsilon_2=-1, \eta_2=-1$
\begin{proposition}
\label{su2-majorana}
Assume $d$ even.
\begin{enumerate}
\item[(i)]
$SU(2)-$Majorana spinors of ${\text Pin}(1,d-1)$ exist when $d=4,6 \text{ mod. }8$.
\item[(ii)]
$SU(2)-$pseudo-Majorana spinors of ${\text Pin}(1,d-1)$ exist when $d=6,8 \text{ mod. }8$.
\end{enumerate}
\end{proposition}
\medskip

\medskip
\noi
\underline{Majorana-Weyl and $SU(2)-$Majorana-Weyl spinors, $d$ even}
\medskip

\noi
Now, from \eqref{chirality} we observe that the Weyl condition is compatible with the
Majorana or the $SU(2)-$Majorana condition if $n+1=2, 4 \text{ mod. } 4$.
Indeed, in such space-time dimension $\chi=(-1)^{\left[\frac{n+1}{2}\right]} \Gamma_0
\cdots \Gamma_{2n-1}$ and the chirality matrix is real.  
This means in this case that the Weyl spinors $\Psi_L$ and $\Psi_R$ can be taken real
or $SU(2)-$Majorana.
This is possible only when the space-time
dimension  $d=2,6 \text { mod. } 8$. Such  spinor will be called  Majorana-Weyl spinors
($d=2 \text{ mod. } 8$) and $SU(2)-$Majorana-Weyl  ($d=6 \text { mod. }8$).  In fact this is the
dimensions for which $\C_0{}_{1,d-1}$ is not simple see Proposition \ref{simple-real} (iv).

\begin{proposition}
\label{majorana-weyl}Assume $d$ even.
\begin{itemize}
\item[(i)]
Majorana-Weyl spinors of ${\text Pin}(1,d-1)$ exist when $d=2 \text{ mod. 8}$.
\item[(i)]
$SU(2)-$Majorana-Weyl spinors of ${\text Pin}(1,d-1)$ exist when $d=6 \text{ mod. 8}$.
\end{itemize}
\end{proposition}

\medskip
 \framebox{Space-time of odd dimension $d=2n+1$}  \\

When the space-time is odd, the matrices $\Gamma_\mu, \mu=0,\cdots,2n-1$ are the same as the ones for
$d=2n$ together with

\beqa 
\label{gammad}
\Gamma_{2n}=i \chi.
\eeqa

\noi
In this case 
taking the matrices \eqref{Gamma(1,d-1)} there is no matrix $U$ such that $U\Gamma_\mu U^{-1}= -\Gamma_\mu$,
because  when $d$ is odd  Clifford algebras are not simple and $\gamma(\epsilon)$ is a number.
This means in particular that differently to the case where $d$ is even we will have
either $B \Gamma_\mu B^{-1}= -\Gamma^\star_\mu$ or $B \Gamma_\mu B^{-1}= \Gamma^\star_\mu$  where the
sign depends on the dimension (similar property holds for the $C$ matrix).
The analogous of \eqref{ABC} writes:

\beqa
\label{ABC2}
\begin{array}{ll}
A \Gamma_\mu A^{-1}=\Gamma_\mu^\dag,& A=\Gamma_0 \cr
B \Gamma_\mu B^{-1}=(-1)^{n}\Gamma_\mu^\star,& B=\Gamma_2 \Gamma_4 \cdots
\Gamma_{2n}\cr
C \Gamma_\mu C^{-1}=(-1)^{n }\Gamma_\mu^t,& C=\Gamma_0 \Gamma_2 \cdots
\Gamma_{2n}
\end{array}
\eeqa

\noi
with 
\beqa
\label{B2}
B B^{*} = (-1)^{\frac{n(n-1)}{2}}.
\eeqa

\noi 
One can check that the matrices $B'=\Gamma_0 \Gamma_1 \cdots \Gamma_{2n-1}$ and
$C'=\Gamma_1 \Gamma_3 \cdots \Gamma_{2n-1}$ give the same relations \eqref{ABC2} as the matrices
$B,C$. As for the even space-time dimension, we introduce
$\epsilon=(-1)^{\frac{n(n-1)}{2}}, \eta=(-1)^n$ which gives

\beqa
\label{signs2}
\begin{array}{lll}
d=1 \text{ mod. }8& \epsilon=+ &\eta=+ \cr
d=3 \text{ mod. }8& \epsilon=+ &\eta=- \cr
d=5 \text{ mod. }8& \epsilon=- &\eta=+ \cr
d=7 \text{ mod. }8& \epsilon=- &\eta=-.
\end{array}
\eeqa

\noi
Then as for even dimensional  space-time  we have the following:

\begin{proposition}
\label{odd}
Assume $d$ is odd. 
\begin{enumerate}
\item[(i)] Majorana spinors of $\text{Pin}(1,d-1)$ exist when $d=1 \text{ mod. }8$. 
\item[(ii)] Pseudo-Majorana spinors of $\text{Pin}(1,d-1)$ exist when $d=3 \text{ mod. }8$
\item[(iii)] $SU(2)-$ Majorana spinors of $\text{Pin}(1,d-1)$ exist when $d=5 \text{ mod. }8$. 
\item[(iv)] $SU(2)-$ pseudo-Majorana spinors of $\text{Pin}(1,d-1)$ exist when 
$d=7 \text{ mod. }8$. 
\end{enumerate}
\end{proposition}

\noi
We conclude this subsection by the following table.

\begin{table}[!ht]
{\small
$$\vbox{\offinterlineskip \halign{
\tv# & \cc{#} & \tv# & \cc{#}  & \tv# &
\cc{#} & \tv# 
&\cc{#} & \tv# 
&\cc{#} & \tv# 
&\cc{#} & \tv# 
&\cc{#} & \tv# 
&\cc{#} & \tv# 
&\cc{#} & \tv# 
&\cc{#} & \tv# 
&\cc{#} & \tv# 
&\cc{#} & \tv# 
&\cc{#} & \tv# 
&\cc{#} & \tv# 
&\cc{#} & \tv# 
 \cr
\noalign{\hrule}
&\cc{$d$}&&$1$ &
&$2$ &
&$3$ &
&$4$ &
&$5$ &
&$6$ &
&$7$ &
&$8$ &
&$9$ &
&$10$ &
&$11$ &
\cr
\noalign{\hrule}
&\cc{M}&
&yes &
&yes &
& &
& &
& &
& &
& &
&yes &
&yes &
&yes &
& &
\cr
&\cc{PM}&
& &
&yes &
&yes &
&yes&
& &
& &
& &
& &
& &
&yes &
&yes & 
\cr
&\cc{MW}&
&&
&yes &
& &
& &
&&
& &
& &
& &
& &
&yes&
&&
\cr
&\cc{SM}&
& &
& &
& &
&yes &
&yes &
&yes &
&&
& &
& &
& &
&&
\cr
&\cc{SPM}&
& &
& &
& &
& &
& &
&yes &
&yes &
&yes &
&&
& &
&&
\cr
&\cc{SMW}&
& &
& &
& &
& &
& &
&yes &
& &
& &
&&
& &
&&
\cr
\noalign{\hrule} 
&\cc{spinor}&
&R &
&R &
&R&
&C&
&PR&
&PR &
&PR&
&C&
&R&
&R&
&R&
\cr
\noalign{\hrule} 
&\cc{SUSY}&
&M &
&MW &
&PM &
&PM &
&SM &
&SMW  &
&SPM &
&M&
&M &
&MW&
&PM &
\cr
\noalign{\hrule} 
&\cc{d.o.f.}&
&$1$ &
&$1$ &
&$2$&
&$4$&
&$8$&
&$8$  &
&$16$&
&$16$&
&$16$&
& $16$&
&$32$ &
\cr
\noalign{\hrule} 
}}$$
}
\caption{{\small  \underline{Types of spinors in various dimensions}.
We have taken the following notations: M for Majorana,
PM for pseudo-Majorana, MW for Majorana-Weyl, SM for $SU(2)-$Majorana,
 SPM for $SU(2)-$pseudo-Majorana and SMW for $SU(2)-$Majorana-Weyl. 
The representations of $\mathfrak{so}(1,d-1)$ are real (R), pseudo-real (PR) or
complex (C) (see section \ref{R-PR-C}).
Recall that the number of real components (d.o.f. in the table)
for the different types of spinors in  $d$ dimensions is for 
(i) a Dirac spinor  $2^{\left[\frac{d}{2}\right]+1}$ (ii) a Majorana or  pseudo-Majorana spinor  
$2^{\left[\frac{d}{2}\right]}$,
(iii) a Majorana-Weyl spinor   $2^{\left[\frac{d}{2}\right]-1}$  (iv) a $SU(2)-$Majorana or
 $SU(2)-$pseudo-Majorana
spinor  $2^{\left[\frac{d}{2}\right]+1}$ and (v) a $SU(2)-$Majorana-Weyl spinor 
 $2^{\left[\frac{d}{2}\right]}$. SUSY means the type of spinors we take to construct a supersymmetric
theory (see section \ref{susy}).
 We note finally that if some kind of spinors
exist for $\text{Pin}(1,d-1)$, they also exist for $\text{Pin}(0,d-2)$ for instance $\text{Pin}(0,8)$
admits a Majorana-Weyl spinors like $\text{Pin}(1,9)$.
 }}\label{tab3}
\end{table}

\noi

\subsection{Real, pseudo-real and complex representations of $\mathfrak{so}(1,d-1)$}
\label{R-PR-C}
Now a group theory touch can be given to the different type of spinors.
As we have seen, we have the following inclusion of algebras:

\beqa
\label{rotation}
\mathfrak{so}(1,d-1)  \subset \C_{1,d-1}.
\eeqa

\noi
Recall that the generators of $\mathfrak{so}(1,d-1)$ are given by 
$\Gamma_{\mu \nu} = \frac14(\Gamma_\mu\Gamma_\nu -\Gamma_\nu\Gamma_\mu), \ \mu, \nu=0,\cdots,d-1$,


\noi 
and  the representation $\Psi_D$ and $B^{-1} \Psi_D^\star$ are equivalent.
As we have seen they can be equated if $B B^\star=1$.  
The representation of $\mathfrak{so}(1,d-1)$
is called real if they can be equated and pseudo-real when they cannot.
In odd dimensional space-time the representations can be either real or pseudo-real. In even dimension in addition to
(pseudo-)real representations  there exist complex representations.

\begin{itemize}
\item[]
When $d=2n+1$, the representation of $\mathfrak{so}(1,2n)$ are real when $d=1, 3 \text{ mod. }8$ and
pseudo-real when $d=5, 7 \text{ mod. }8$.
\item[]
When  $d=2n$, we introduce 

\beqa
\label{LR}
\Sigma_{\mu \nu}^{\pm}= \frac12(1 \pm \chi) \Gamma_{\mu \nu}
\eeqa

\noi which are the generators of the spinor representations ${\cal S}_\mp$. From \eqref{chirality}
we get $\chi^\star= (-1)^{n+1} B\chi B^{-1},$ thus

\beqa
\label{conj-even}
\Sigma^\pm_{\mu \nu}{}^\star = \left\{
\begin{array}{ll}
B \Sigma^\pm_{\mu \nu} B^{-1}&\text{when } n \text{ odd} \cr
B \Sigma^\mp_{\mu \nu} B^{-1} &\text{when } n \text{ even}.
\end{array}\right.
\eeqa

\noi
This means that the complex conjugate of ${\cal S}_\pm$ is equal to itself when $d=4k+2$. 
The representation
will be real when $d=2 \text{ mod. } 8$ ($B B^\star =1$) and pseudo-real when $d=6 \text { mod. } 8$
 ($B B^\star =-1$). However, 
when $d=4k$ the complex conjugate of ${\cal S}_\pm$ is ${\cal S}_\mp$
and the representation is complex.
\end{itemize}

\begin{remark}
The results above can also give some insight on the structure of  Clifford
algebras. When $d$ is even, there always exists a matrix $B$ such that $B \Gamma_\mu B^{-1}= \Gamma_\mu$
(see \eqref{ABC}).
If $BB^\star=1$ the Clifford algebra is real $(d=0,2 \text { mod. } 8$) and if
$BB^\star=-1$ the algebra is quaternionian ($d=4,6 \text { mod. } 8$).
(Notice that for $d=4$ the pseudo-Majorana spinors a taken with the oposite choise {\it i.e}
with $B$ s.t. $B \Gamma_\mu B^{-1}=-\Gamma^\star_\mu$.) When $d$ is odd,
either $\e^2=1$ ($\rho(\e)=\pm$) or $\e^2=-1$ ($\rho(\e)=\pm i$). When $\e^2=1$ (or when $B \Gamma_\mu B^{-1}= \Gamma_\mu$) 
the Clifford algebra is real when $BB^\star=1$ ($d=1 \text{ mod. } 8$) and quaternionian when 
$BB^\star=-1$ ($d=5 \text{ mod. } 8$)
although when   $\e^2=-1$ (or when $B \Gamma_\mu B^{-1}= -\Gamma_\mu$) the Clifford algebra is complex
$(d=3,7 \text{ mod. } 8$). This can be compared with table \ref{tab2}.
\end{remark}

\subsection{Properties of (anti-)symmetry of the $\Gamma-$matrices}
The matrices

\beqa
\label{covariant-Dirac}
\Gamma^{(\ell)}_{\mu_1 \ldots \mu_\ell} = \frac{1}{\ell !}
\sum \limits_{\sigma \in \Sigma_{\ell}} \epsilon(\sigma) \Gamma_{\mu_{\sigma(1)} \ldots \mu_{\sigma(\ell)}},
\eeqa

\noi
with $\ell =0, \cdots, d$ when $d$ is even and with $\ell =0 \cdots, \frac{d-1}{2}$ when $d$ is
odd constitute a basis of the representation of $\C_{1,d-1}$. 
(In $d$ dimensions we have $\begin{pmatrix}d \\ \ell\end{pmatrix}$ independant matrices $\Gamma^\ell_{\mu_1\cdots \mu_\ell}$
and
  $2^{2n} = \sum \limits_{\ell=0}^{2n}\begin{pmatrix}2n \\ \ell \end{pmatrix}$  when $d=2n$ or
  $2^{2n} = \sum \limits_{\ell=0}^{n}\begin{pmatrix}2n+1 \\\ell \end{pmatrix}$ when $d=2n+1$.)
Note the normalisation factor
$\Gamma_{\mu \nu} = \frac12 \Gamma^{(2)}_{\mu \nu}.$ Now using the matrices $C_1$ and $C_2$ in
\eqref{ABC} and the matrix $C$ in \eqref{ABC2}, on can show that the matrices 
$\Gamma^{(\ell)} C^{-1}$ are either fully symmetric or fully antisymmetric. It is just a matter of a simple 
calculation to check the following:

\beqa
\label{sym}
\begin{array}{ll}
(\Gamma_{\mu_1 \cdots \mu_\ell}^{(\ell)} C_1^{-1})^t=(-1)^{\frac12\left(\ell + n)^2+(\ell-n)\right)}
\Gamma_{\mu_1 \cdots \mu_\ell}^{(\ell)} C_1^{-1}&d=2n \cr
(\Gamma_{\mu_1 \cdots \mu_\ell}^{(\ell)} C_2^{-1})^t=(-1)^{\frac12\left(\ell + n)^2+(n-\ell)\right)
}\Gamma_{\mu_1 \cdots \mu_\ell}^{(\ell)} C_2^{-1}&d=2n \cr
(\Gamma_{\mu_1 \cdots \mu_\ell}^{(\ell)} C^{-1})^t=(-1)^{\frac12\left(\ell + n)^2+(n-\ell)\right)}
\Gamma_{\mu_1 \cdots \mu_\ell}^{(\ell)} C^{-1}&d=2n+1 
\end{array}
\eeqa

\noi
We now summarize in the following table the symmetry properties of the $\Gamma C^{-1}-$matrices. \\

\begin{table}[!ht]
{\small
$$\vbox{\offinterlineskip \halign{
\tv# & \cc{#} & \tv# & \cc{#}  & \tv# &
\cc{#} & \tv# 
&\cc{#} & \tv# 
&\cc{#} & \tv# 
&\cc{#} & \tv# 
&\cc{#} & \tv# 
&\cc{#} & \tv# 
&\cc{#} & \tv# 
&\cc{#} & \tv# 
&\cc{#} & \tv# 
&\cc{#} & \tv# 
&\cc{#} & \tv# 
&\cc{#} & \tv# 
&\cc{#} & \tv# 
&\cc{#} & \tv# 
 \cr
\noalign{\hrule}
&\cc{$$}&
&$0$ &
&$1$ &
&$2$ &
&$3$ &
&$4$ &
&$5$ &
&$6$ &
&$7$ &
&$8$ &
&$9$ &
&$10$ &
\cr
\noalign{\hrule}
&$1$&
&S & 
& & 
& &
& &
& &
& &
& &
& & 
& &
&  &
& &
\cr
&$2$&
&S &
&S &
&A &
& &
& &
& &
& &
& &
& &
& &
&  & 
\cr
&$2$&
&A &
&S &
&S &
& &
& &
& &
& &
& &
& &
& &
&  & 
\cr
&$3$&
&A &
&S &
& &
& &
& &
& &
& &
& &
& &
& &
&  & 
\cr
&$4$&
&A &
&S &
&S &
&A &
&A &
& &
& &
& &
& &
& &
&  & 
\cr
&$4$&
&A &
&A &
&S &
&S &
&A &
& &
& &
& &
& &
& &
&  & 
\cr
&$5$&
&A &
&A &
&S &
& &
& &
& &
& &
& &
& &
& &
&  & 
\cr
&$6$&
&A &
&A &
&S &
&S &
&A &
&A &
&S&
& &
& &
& &
&  & 
\cr
&$6$&
&S &
&A &
&A &
&S &
&S &
&A &
&A &
& &
& &
& &
&  & 
\cr
&$7$&
&S &
&A &
&A &
&S &
& &
& &
& &
& &
& &
& &
&  & 
\cr
&$8$&
&S &
&A &
&A &
&S &
&S &
&A &
&A&
&S &
&S &
& &
&  & 
\cr
&$8$&
&S &
&S &
&A &
&A &
&S &
&S &
&A&
&A &
&S &
& &
&  & 
\cr
&$9$&
&S &
&S &
&A &
&A &
&S &
& &
& &
& &
& &
& &
&  & 
\cr
&$10$&
&S &
&S &
&A &
&A &
&S &
&S &
&A&
&A &
&S &
&S &
& A & 
\cr
&$10$&
&A &
&S &
&S &
&A &
&A &
&S &
&S&
&A &
&A &
&S &
& S & 
\cr
&$11$&
&A &
&S &
&S &
&A &
&A &
&S &
& &
& &
& &
& &
&  & 
\cr 
\noalign{\hrule} 
}}$$
}
\caption{{\small \underline{Symmetry of $\Gamma^{(\ell)}C^{-1}$}. 
The first row  indicates the type of matrix $\ell$  for 
$\Gamma^{(\ell)} C^{-1}$ and the first column  the space-time  dimension. 
When $d$ is even
we have two series of matrices $\Gamma^{(\ell)} C_1^{-1}$ and $\Gamma^{(\ell)} C_2^{-1}$
respectively in the first and second line. Finally, 
S  (resp. A) indicates that $\Gamma^{(\ell)} C^{-1}$ is symmetric (resp. antisymmetric).
 }}\label{tab4}
\end{table}

Now we would like to give some duality properties. We define 
$\varepsilon^{\mu_1 \cdots \mu_{2n}}$ the Levi-Civita tensor (equal to 
the signature of the permutation that brings $\mu_1 \cdots \mu_{2n}$ to $0,1,\cdots,2n-1,
\varepsilon^{0 \cdots 2n-1}=1$). We also introduce  $\Gamma^\mu : \Gamma_\mu = 
\eta_{\mu \nu} \Gamma^\nu$ 
and $\varepsilon_{\mu_1 \cdots \mu_{2n}}=\varepsilon^{\nu_1 \cdots \nu_{2n}} \eta_{\mu_1 \nu_1}
\cdots \eta_{\mu_{2n} \nu_{2n}}$ $\varepsilon_{0 \cdots 2n-1}=-1$) 
and $\Gamma= \Gamma_0 \dots \Gamma_{2n-1}$. From,

\beqa
\label{dual}
\Gamma^{(\ell)}{}^{\mu_1 \cdots \mu_{2n}} \Gamma&=& (-1)^n \varepsilon^{\mu_1\cdots \mu_{2n}} \\
\Gamma^{(\ell)}{}^{\mu_1 \cdots \mu_\ell} \Gamma &=& \frac{(-1)^{\frac{\ell(\ell-1)}{2}}}{(2n -\ell) !}
 \varepsilon^{\mu_1\cdots \mu_{2n}} \Gamma^{(\ell)}{}_{\mu_{\ell +1} \cdots \mu_{2n}} \nonumber
\eeqa

\noi
we get

\beqa
\label{dual2}
\begin{array}{ll}
{}^\star \left(\Gamma^{(n)} \pm i \Gamma^{(n)} \Gamma \right)= \pm i (-1)^{\frac{n(n+1)}{2}+1}
\left(\Gamma^{(n)} \pm i \Gamma^{(n)} \Gamma \right) & \text{ when } n \text{ is even} \cr
{}^\star \left(\Gamma^{(n)} \pm  \Gamma^{(n)} \Gamma \right)= \pm  (-1)^{\frac{n(n+1)}{2}+1}
\left(\Gamma^{(n)} \pm  \Gamma^{(n)} \Gamma \right) & \text{ when } n \text{ is odd}
\end{array}
\eeqa

\noi
with ${}^\star$ the Hodge dual
$({}^\star X_{\mu_1 \cdots \mu_p} = \frac{1}{(2n-p)!} \varepsilon_{\mu_1 \cdots \mu_p \nu_1 \cdots \nu_{2n-p}} 
X^{ \nu_1 \cdots \nu_{2n-p}}$).
Thus this means that the above matrices  are (anti-)self-dual.

\subsection{Product of spinors}
Now we give the decomposition of the product of spinor representations.
We recall that $\text{End}({\cal S}) \cong {\cal S} \otimes {\cal S}^\star$. Furthermore, 
given $\Psi_D$ and $\Xi_D$ two Dirac spinors, recall 

\beqa
\label{PS}
S=\bar \Xi_D\Psi_D = \Xi^\dag_D \Gamma_0 \Psi_D
\eeqa

\noi
define  invariant scalar products (see \eqref{equiv}). 

\begin{remark}
\label{adjoint}
When the signature is $(t,s)$ we take $\Gamma_1\cdots \Gamma_t$ instead of $\Gamma_0$
to set  $  \bar \Psi = \Psi^\dag \Gamma_1\cdots \Gamma_t $.
This legitimate \eqref{beta}.
\end{remark}

\noi
Similarly,

\beqa
\label{p-form}
\bar \Xi \Gamma^{(\ell)}_{\mu_1 \cdots \mu_\ell} \Psi_D
\eeqa

\noi
transforms as an $\ell^{\text{th}}$ order antisymmetric tensor.
Now, from \eqref{equiv} it is easy to see that 

\beqa
\label{p-form2}
T_{\mu_1\cdots \mu_\ell}= \Xi^t C \Gamma^{(\ell)}_{\mu_1 \cdots \mu_\ell} \Psi_D
\eeqa

\noi
transforms also as an $\ell^{\text{th}}$ order antisymmetric tensor.
Now, since the matrices $\Gamma^{(\ell)}_{\mu_1\cdots \mu_\ell}$
with $\ell =0,\cdots d-1$ when $d$ is even and $\ell=0,\cdots \frac{d-1}{2}$ when
$d$ is odd constitute a basis
of ${\cal M}_{2^{\left[\frac{d}{2}\right]}}(\CC)$ and since 
$\Xi^t C \in {\cal S}^\star$ (the dual space of ${\cal S}$)
we have 
 
\beqa 
\label{tensor} 
{\cal S} \otimes {\cal S}^\star = \left\{ 
\begin{array}{ll} 
\CC \oplus E \oplus  \Lambda^2(E)  \oplus \cdots \oplus \Lambda^d(E)&
\text{when }d \text{ is even} \cr
 \CC \oplus E \oplus  \Lambda^3(E)  \oplus \cdots \oplus 
\Lambda^{\left[\frac{d}{2}\right]}(E)&
\text{when }d \text{ is odd}.
\end{array}
\right.
\eeqa



\noi This decomposition has to be compared with Remark
\ref{exterior}.

Now, when $d=2n$ is even one can calculate the product of Weyl spinors. Let
$\Psi_{\epsilon_1} \in {\cal S}_{\epsilon_1}, \epsilon_1=\pm$  and $\Xi^t_{\epsilon_2} C
\in {\cal S}^\star_{\epsilon_2}, \epsilon_2=\pm$ be two Weyl spinors. Recall that we have
$\chi \Psi_{\epsilon_1} = -\epsilon_1 \Psi_{\epsilon_1}$. Now, using
$C \chi C^{-1}=  \Gamma_0^t \cdots \Gamma_{d-1}^t$. By \eqref{conj} the RHS writes
$(-1)^n \chi$ but rearranging the product the RHS also writes $(-1)^n\chi^t$. Thus

\beqa 
\label{chi-C} 
\chi^t = \chi, \ \ C \chi=(-1)^n \chi C, 
\eeqa

\noi and we have

\beqa
\Xi^t_{\epsilon_2} C \Gamma_{\mu_1 \cdots \mu_\ell}^{(\ell)} \Psi_{\epsilon_1}&=&
\epsilon_1 \Xi^t_{\epsilon_2} C \Gamma_{\mu_1 \cdots \mu_\ell}^{(\ell)} \chi \Psi_{\epsilon_1}=
\epsilon_1 (-1)^\ell \Xi^t_{\epsilon_2} C \chi\Gamma_{\mu_1 \cdots \mu_\ell}^{(\ell)}  \Psi_{\epsilon_1}
\nonumber \\
&=&
\epsilon_1 (-1)^{n + \ell} \Xi_{\epsilon_2}^t\chi C\Gamma_{\mu_1 \cdots \mu_\ell}^{(\ell)}  
\Psi_{\epsilon_1}=
\epsilon_1 (-1)^{n + \ell} (\chi \Xi_{\epsilon_2})^t C\Gamma_{\mu_1 \cdots \mu_\ell}^{(\ell)}  
\Psi_{\epsilon_1} \nonumber \\
&=& \epsilon_1 \epsilon_2(-1)^{n + \ell} \Xi_{\epsilon_2}^t C\Gamma_{\mu_1 \cdots \mu_\ell}^{(\ell)}  
\Psi_{\epsilon_1}
\eeqa

\noi
and  $\Xi^t_{\epsilon_2} C \Gamma_{\mu_1 \cdots \mu_\ell}^{(\ell)} \Psi_{\epsilon_1}=0$
if $ \epsilon_1 \epsilon_2(-1)^{n + \ell}=-1 $. This finally gives

\beqa
\label{tensor2}
{\cal S}_+ \otimes  {\cal S}_+^\star = 
\left\{ \begin{array}{ll}
\CC \oplus \Lambda^2(E) \oplus \cdots \oplus \Lambda^n(E)_+& \text{ when } n \text{ is even} \cr
E \oplus  \Lambda(E) \oplus\cdots \oplus \Lambda^{n}(E)_+& \text{ when } n \text{ is odd} 
\end{array} \right. \nonumber \\
\\
{\cal S}_+ \otimes  {\cal S}^\star_- = 
\left\{ \begin{array}{ll}
E \oplus  \Lambda^3(E) \oplus\cdots \oplus \Lambda^{n-1}(E)& \text{ when } n \text{ is even} \cr
\CC \oplus \Lambda^2(E) \oplus \cdots \oplus\Lambda^{n-1}(E)& \text{ when } n \text{ is odd} 
\end{array} \right. .\nonumber
\eeqa

\noi
A special attention has to be paid for the antisymmetric tensors of order $n$. Indeed, 
the antisymmetric tensor of order $n$ is a reducible representation of $SO(1,2n-1)$ and 
can be decomposed into self-dual and anti-self-dual tensors (see \eqref{dual2}) (denoted $ \Lambda^n(E)_\pm$):

\beqa
\Lambda^n(E) = \Lambda^n(E)_+\oplus \Lambda^n(E)_-.
\eeqa

\noi
Now a simple counting of the dimensions on both sides in \eqref{tensor2} shows that
only the self- (or anti-self-) dual $n^{\text {th}}-$order tensors appears in the decomposition.
(The dimension of $\Lambda^p(E)$ is $\begin{pmatrix} 2n \\ p \end{pmatrix}$  and of  $\Lambda^n(E)_\pm$ is 
$\frac12\begin{pmatrix} 2n \\ n \end{pmatrix}$.)
The choice  $\Lambda^n(E)_+$ is a matter of convention.

If we rewrite explictly these relation ({\it e.g.} \eqref{tensor} for $d$ even), we have

\beqa
\label{tensor-p-formes} 
\Psi_D \otimes  \Xi_D^t C &=& \sum \limits_{\ell =0}^d \frac{1}{\ell!} 
T^{(\ell)}{}^{\mu_1 \cdots \mu_\ell} \Gamma^{(\ell)}_{\mu_1 \cdots \mu_\ell}  \nonumber \\
\Psi_D \otimes  \Xi_D^t  &=& \sum \limits_{\ell =0}^d \frac{1}{\ell!} 
T^{(\ell)}{}^{\mu_1 \cdots \mu_\ell} \Gamma^{(\ell)}_{\mu_1 \cdots \mu_\ell} C^{-1} 
\eeqa

\noi
with $T^{(\ell)}$ an antisymmetric $\ell-$th order tensor.
\section{Clifford algebras and supersymmetry}
\label{susy}
\renewcommand{\theequation}{5.\arabic{equation}}   
\setcounter{equation}{0}
In this section, with the help of the previous sections, we study   
supersymmetric algebras with a special attention to the four, ten and eleven-dimensional
space-times. We then study  the representations of the considered supersymmetric
algebras and show that representation spaces contain an equal number of bosons and
 fermions. Supersymmetry turns out to be a symmetry which mixes non-trivially the
bosons and the fermions since one multiplet contains bosons and fermions together.
We also show how four and ten dimensional supersymmetry are related to eleven dimensional
supersymmetry by compactification or dimensional reduction.
\subsection{Non-trivial extensions of the Poincar\'e algebra}

Describing   the laws of physics in terms of  underlying symmetries has 
always been a powerful tool. 
For instance the Casimir operators of the Poincar\'e algebra \eqref{poincare}
are related to the mass and the spin of elementary particles as the electron
or the photon. Moreover, it has been understood that all the fundamental
interactions (electromagnetic, week and strong interactions) are related to
the Lie algebra $\mathfrak{u}(1)_Y \times \mathfrak{su}(2)_L \times \mathfrak{su}(3)_c$,
in the so-called standard model (see {\it e.g.} \cite{SM} and references therein). The standard model is then 
described by the Lie algebra
$ \mathfrak{iso}(1,3) \times \mathfrak{u}(1)_Y \times 
\mathfrak{su}(2)_L \times \mathfrak{su}(3)_c$, where $\mathfrak{iso}(1,3)$ is related
to space-time symmetries and  $\mathfrak{u}(1)_Y \times 
\mathfrak{su}(2)_L \times \mathfrak{su}(3)_c$ to internal symmetries.
Even if the standard  model is the physical theory were the confrontation
between experimental results and theoretical predictions is in an extremely good
accordance, there is
strong arguments (which cannot be summarized here) that it is not the final theory.
 
Thus, to understand the properties of elementary particles,  it is then interesting to study
the kind of symmetries which are allowed in space-time. Within the
framework of Quantum Field Theory (unitarity of the $S$ matrix {\it etc}.),
S. Coleman and  J. Mandula have  shown  \cite{cm} that if the symmetries are
described in terms of Lie algebras,  only trivial extensions
of the Poincar\'e algebra can be obtained. Namely, the fundamental symmetries are
based on $ \mathfrak{iso}(1,3) \times \mathfrak{g}$
with $  \mathfrak{g}  \supseteq \mathfrak{u}(1)_Y \times \mathfrak{su}(2)_L \times \mathfrak{su}(3)_c$
a compact  Lie algebra describing the fundamental interactions and
$[ \mathfrak{iso}(1,3),\mathfrak{g}]=0$. Several algebras, in relation to
phenomenology, have been investigated (see {\it e.g.} \cite{SM, gut}) such as $\mathfrak{su}(5), \mathfrak{so}(10), 
\mathfrak{e}_6$ 
{\it etc.} Such theories are usually refer to  ``Grand-Unified-Theories'' or  GUT, {\it i.e.}
theories  which unify all the fundamental interactions. The fact that elements of
$\mathfrak{g}$ and $\mathfrak{iso}(1,3)$ commute means that we have a trivial
extension of the Poincar\'e algebra.

Then, R.~Haag, J.~T.~Lopuszanski and M.~F.~Sohnius \cite{hls} understood that is was
possible to extend in a non-trivial way the symmetries of space-time within
the framework of Lie superalgebras (see Definition \ref{superlie}) in an unique
way called supersymmetry. 
We first give the definition of
a Lie superalgebras, and then we show how to construct a supersymmetric theory in
any space-time dimensions using the results established in Section \ref{spinors}.

\begin{definition}
\label{superlie}
A Lie (complex or real) superalgebra is a $\mathbb{Z}_2-$graded vector space 
$\mathfrak{g}=\mathfrak{g}_0 \oplus \mathfrak{g}_1$ endowed with the following structure
\begin{enumerate}
\item $\mathfrak{g}_0$ is a Lie algebra, we denote by $[\ ,\ ]$ the bracket on  $\mathfrak{g}_0$
($\left[\mathfrak{g}_0, \mathfrak{g}_0 \right] \subseteq \mathfrak{g}_0$);
\item $\mathfrak{g}_1$  is a representation of $\mathfrak{g}_0$
($\left[\mathfrak{g}_0, \mathfrak{g}_1 \right] \subseteq \mathfrak{g}_1$);
\item there exits a $\mathfrak{g}_0-$equivariant mapping 
$\left\{ \ ,\  \right\}: S^2\left(\mathfrak{g}_1\right) \to \mathfrak{g}_0$ where 
$S^2\left(\mathfrak{g}_1\right)$ denotes the two-fold symmetric product of $\mathfrak{g}_1$
($\left\{\mathfrak{g}_1, \mathfrak{g}_1 \right\} \subseteq \mathfrak{g}_0$);
\item The following Jacobi identities hold ($\forall \ b_1,b_2,b_3 \in \mathfrak{g}_0, \forall \  f_1,f_2,f_3 \in
 \mathfrak{g}_1$)
\beqa
\label{jacobi}
&&\left[\left[b_1,b_2\right],b_3\right] +
\left[\left[b_2,b_3\right],b_1\right] +
\left[\left[b_3,b_1\right],b_2\right] =0 \nonumber \\
&&\left[\left[b_1,b_2\right],f_3\right] +
\left[\left[b_2,f_3\right],b_1\right] +
\left[\left[f_3,b_1\right],b_2\right]  =0  \\
&&\left[b_1,\left\{f_2,f_3\right\}\right] -
\left\{\left[b_1,f_2 \right],f_3\right\}  -
\left\{f_2,\left[b_1,f_3\right] \right\} =0 \nonumber \\
&& \left[ f_1,\left\{f_2,f_{3}\right\} \right] + \left[ f_2,\left\{f_3,f_{1}\right\} \right] 
 + \left[ f_3,\left\{f_1,f_{2}\right\} \right]  =0.
 \nonumber
\eeqa
\end{enumerate}
The generators of zero  (resp. one) gradation are called the bosonic (resp. fermionic)  generators or
$\mathfrak{g}_0$ (resp. $\mathfrak{g}_1$) is called the bosonic (resp.  fermionic) part of the Lie
superalgebra. The first Jacobi identity is the usual Jacobi identity for Lie algebras, the second says 
that $\mathfrak{g}_1$ is a representation of  $\mathfrak{g}_0$, the third identity is the equivariance
of $\{ \  ,\  \}$. These identities are just consequences of  
$1., 2.$ and $3.$ respectively. However, the fourth Jacobi identity
which is an extra constraint, is just the $\mathbb Z_2-$graded Leibniz rule.
\end{definition}

\noi
The supersymmetric extension of the Poincar\'e algebra is constructed,
in the framework of Lie superalgebras, by adjoining to the 
Poincar\'e generators anticommuting elements, called supercharges
(we denote $Q$), which belong to the spinor
representation of the Poincar\'e algebra. Thus the supersymmetric algebra is
a Lie superalgebra  $\mathfrak{g}= \mathfrak{iso}(1,d-1) \oplus {\cal S}$
with  brackets

\beqa
[L,L] =L, \ [L,P]=P,\  [L,Q]=Q,\ [P,Q]=0,\ \{Q,Q\}=P,
\eeqa

\noi
with $(L,P)$ the generators of the Poincar\'e algebra that belong to the bosonic part of the algebra
and $Q$  the fermionic part of the algebra.
This extension is non-trivial, because the supercharges $Q$ are spinors, and thus 
do not commute with the generators of the Lorentz algebra.
However, the precise definition depends on the space-time 
dimension because the reality properties
of the spinor charges  and  hence the structure of the algebra depends
 on the dimensions (see table \ref{tab3}). A systematic study of  supersymmetric extensions
has been undertaken in \cite{s}. Table \ref{tab3}  indicates the type of spinor
we take in  various dimensions. Furthermore, the number of supercharges  $N$ we
consider is such that the total spinorial degrees of freedom is less than $32$ (see section \ref{repsusy4d},
Remark \ref{N<8}), 
This gives the possible choices for the spinorial charges
(see Table \ref{tab3})

\beqa
\label{susy-d}
\begin{array}{lll}
d=4&N\ \text{pseudo-Majorana}&1< N \le 8 \cr
d=5&N\ SU(2)\text{-Majorana}&1< N \le 4 \cr
d=6&(N_+,N_-)\ \text{ (right-, left-) handed }  \cr
&SU(2)\text{-Majorana-Weyl}& 1<N_+ + N_- \le 4  \cr
d=7&N \ SU(2)\text{-pseudo-Majorana}& 1<N \le 2  \cr
d=8& N\ \text{Majorana}& 1<N \le 2 \cr
d=9&N \ \text{Majorana}& 1<N \le 2 \cr
d=10&(N_+,N_-)\ \text{ (right-, left-) handed }  \cr
&\text{ Majorana-Weyl}& 1<N_+ + N_- \le 2  \cr
d=11&N \ \text{pseudo-Majorana}& N = 1  \cr
\end{array}
\eeqa

\subsection{Algebra of supersymmetry in various dimensions}
Using  \eqref{susy-d} we give the  supersymmetric algebras in
four, ten and eleven dimensions. Supersymmetry in other space-time dimensions are constructed
along the same lines \cite{s}. 
In this section, we will not give
 precise references of the subject, one may for instances see \cite{west} and references
therein (in particular we will not refer to the original papers on the
subject\footnote{Supersymmerty and supergravity is an intense subject of research. If one goes
to the particle physics data basis http://www.slac.standford.edu and types {\it find
title supersymmetry or sypergravity}, one has 8767 different publications 
or types {\it find k supersymmetry or supergravity}, one has 42449 answers (the 1.06.2005) }).

\subsubsection{Supersymmetry in four dimensions}
Now, we give the precise structure of the supersymmetric extension of the Poincar\'e
algebra in  four dimensions. 
 A standard reference on the subject is \cite{wb} (see also \cite{so,west}).
The bosonic (or even) part of the algebra is given by the Poincar\'e  generators $L_{\mu \nu}, P_\mu$. 
The fermionic (or odd) sector  is constituted of $N$ Majorana  spinor supercharges $Q_I, I=1,\cdots,N$
(see table \ref{tab3}).
The  algebraic structure  is given by three types of bracket:
(i) $[\text {even }, \text{ even}]$, (ii) $[\text {even }, \text{ odd}]$ and 
(iii) $\{\text {odd }, \text{ odd}\}$, where even/odd  means bosonic/fermionic generators.
The first types of bracket is the Poincar\'e algebra \eqref{poincare}.
The action of the Poincar\'e algebra onto the fermionic supercharges is given by

\beqa
\label{susy4d2}
[L_{\mu \nu}, Q_I]= \Gamma_{\mu \nu} Q_I, \ \ [P_\mu,Q_I]=0,
\eeqa

\noi
this is the second type of bracket.
The first equation is due to the fact that $Q$ is in the spinor representation
of the Lorentz algebra
(with matrix representation $\Gamma_{\mu \nu}$  \eqref{spin-generators}). Since $P_\mu$ transforms like a vector,
and among the $\Gamma-$matrices only $\Gamma_\mu$ transforms as a vector 
(see \eqref{p-form1}),  the second relation
would be  $[P_\mu,Q]=c \Gamma_\mu Q$. But the Jacobi
identity \eqref{jacobi} involving $(P,P,Q)$ leads to $c=0$.

Now, we study the last type of brackets involving only odd generators.
To be compatible with the literature \cite{wb,so}, we will not use the Dirac $\Gamma-$matrices 
given in Section \ref{spinors}, but 

\beqa
\label{gamma4}
\Gamma_\mu=\begin{pmatrix}0& \sigma_\mu \\ \bar \sigma_\mu &0 \end{pmatrix}
\eeqa

\noi
with $\sigma_\mu$ the Pauli matrices given in \eqref{pauli}  
and $(\bar \sigma_0, \bar \sigma_i)=
(\sigma_0, - \sigma_i)$. In this representation the chirality matrix reads
$\chi_4= \begin{pmatrix}\sigma_0& 0 \\ 0&-\sigma_0 \end{pmatrix}$ and 
the $B_4$, $C_4$ matrices \eqref{ABC} reduce to $B_4=i \Gamma_2, \  $ $C_4=i\Gamma_0 \Gamma_2$.   
The Majorana spinor  supercharges are defined  by

\beqa 
\label{majorana4}
Q_I= \begin{pmatrix} Q_L{}_I \\ Q_R{}_I \end{pmatrix}
\eeqa

\noi 
with $Q_L{}_I$  (resp. $Q_R{}_I$)   complex left-handed  (resp. right-handed) Weyl spinors. 
The condition $Q_I^\star= B_4 Q_I$ gives
 $Q^\star_{RI}= -i \sigma_2 Q_L{}_I$ and the complex conjugate of a right-handed spinor
is a left-handed spinor as we have seen in section \ref{R-PR-C}. Finally, 
remember that in four dimensions we have for the product of two
spinors (see \eqref{tensor-p-formes}) 
$
\Psi \otimes  \Xi^t  =  T^{(0)} C_4^{-1} +  T^{(1)}{}^\mu \Gamma_\mu C_4^{-1} +
\frac12  T^{(2)}{}^{\mu \nu} \Gamma_{\mu \nu} C_4^{-1} +
\frac16  T^{(3)}{}^{\mu \nu \rho} \Gamma_{\mu \nu \rho } C_4^{-1} +
\frac{1}{24}  T^{(4)}{}^{\mu \nu \rho \sigma} \Gamma_{\mu \nu \rho  \sigma} C_4^{-1}$.\\

To construct the last type of brackets, we make the following remarks:
\begin{enumerate}
\item the fermionic part of the algebra as to close onto the bosonic part, which 
 reduces to the Lorentz generators $L_{\mu \nu}$
and to the generators of the space-time translations $P_\mu$;
\item the only symmetric Dirac $\Gamma-$ matrices are $\Gamma_\mu$ and $\Gamma_{\mu \nu}$
(see table \ref{tab4} and equation \eqref{tensor-p-formes});
\item the Jacobi identity \eqref{jacobi}  as to be satisfied.
\end{enumerate}

\noi
The points 1 and 2 give

$$ \left\{Q_I, Q_J^t\right\}= \delta_{IJ} (a P^\mu \Gamma_\mu C^{-1}_4 +b L^{\mu \nu} \Gamma_{\mu \nu}C^{-1}_4)$$

\noi
with $Q_J^t$ the transpose of $Q_J$
(in fact instead of $\delta_{IJ}$ we could have obtained a symmetric second order tensor, which can always
be diagonalised \cite{wb}.) Now, the Jacobi identity involving $(Q,Q,P)$ gives $b=0$ since $L_{\mu \nu}$ 
and $P_\mu$
do not commute. Finally for conventional reason, we chose $a=-2$. 

Now we observe that
the bosonic part of the algebra can be enlarged  by introducing
some new (real) scalar generators $X_{IJ}, Y_{IJ}$ commuting  with all the bosonic elements and being 
antisymmetric  $X_{IJ}=-X_{JI},Y_{IJ}=-Y_{JI}$ \cite{wb} (the new generators are called central charges). Since,
the matrices $C^{-1}_4$ and $i \chi C^{-1}_4$ are antisymmetric (see table \ref{tab4})  the algebra 
extends to

\beqa
\label{susy4d}
 \left\{Q_I, Q^t_J\right\}= -2 \delta_{IJ} P^\mu \Gamma_\mu C^{-1}_4  + X_{IJ} C^{-1}_4 +i Y_{IJ} \chi C^{-1}_4.
\eeqa

\noi
Now, studying the various Jacobi identities involving $X,Y$ we can show  \cite{wb}

\beqa
\label{Z}
[X_{IJ}, \text{ anything }]=0, \ \ [Y_{IJ}, \text{ anything }]=0.
\eeqa

 \noi
The Lie superalgebra defined by \eqref{poincare},  \eqref{susy4d2}, \eqref{susy4d},  and \eqref{Z} 
 is called the four-dimensional $N-$extended super-Poincar\'e
algebra.

\begin{remark}
\label{auto}
Since, we have $N$ copies of the (complex) supercharges $Q_L{}_I$ and $Q_R{}_I$, 
this allows the action to the 
automorphism group for which $Q_I{}_L$ are in the $N-$dimensional representation of 
$G \subseteq U(N)$ and  $Q_R{}_I$ is
in the complex conjugate representation. The full $N-$extended superalgebra as to be supplemented with
$[T_a, Q_L{}_I]= (t_a)_I{}^J Q_L{}_J, \ \ [T_a, Q_R{}_I]= (t_a)^\star{}_I{}^J Q_R{}_J$ with $T_a$ the
generators of $\mathfrak{g}$ (the Lie algebra of $G$) and $t_a$ the 
$N \times N$ matrices corresponding to the 
$N-$dimensional 
representation of $\mathfrak{g}$ and  $(t_a)^\star$ the matrices of the complex conjugate representation \cite{wb}.
With $Q_I=\begin{pmatrix} Q_{LI} \\ Q_{RI}\end{pmatrix}$ these relations become
$\left[T_a,Q_I\right]=  
\frac{ (t_a){}_I{}^J + (t^*_a){}_I{}^J}{2} Q_J + i \chi 
\frac{i \left((t_a){}_I{}^J - (t^*_a){}_I{}^J\right)}{2} Q_J$.
\end{remark}

\begin{remark}
\label{2-4}
The algebra was presented in a formalism where the spinors have four components. 
This algebra can also be realized in the 
two components notations, with the Weyl spinors $Q_L$ and $Q_R$. From

\beqa
\left\{ {  Q}_I, { Q}_J^t \right\}= \left\{\begin{pmatrix} Q_L{}_I \\ Q_R{}_I \end{pmatrix},
\begin{pmatrix} Q^t_{LJ}& Q^t_{RJ} \end{pmatrix} \right\}=
\begin{pmatrix}
\left\{Q_L{}_I,  Q^t_{LJ}\right\} & \left\{ Q_L{}_I,  Q^t_{RJ}\right\} \\
\left\{Q_R{}_I,  Q^t_{LJ}\right\} & \left\{Q_R{}_I,   Q^t_{RJ}\right\}
\end{pmatrix}.
\eeqa

\noi
the algebra \eqref{susy4d} reduces to

\beqa
 \left\{Q_L{}_I,Q^t_{LJ}\right\}=i(X_{IJ} + i Y_{IJ})\sigma_2, \ \  \left\{Q_L{}_I,Q^t_{RJ}\right\}=2i 
\delta_{IJ}P^\mu  \sigma_\mu \sigma_2.
\nonumber 
\eeqa
\noi
These  brackets could also have been deduced from \eqref{tensor2}.
\end{remark}

\noi 
The study of representation of the supersymmetric algebra (see sect \ref{repsusy4d}) will in fact give 
$N \le 8$. This means that the maximum number of fermionic degrees of freedom is $32 \ (=4\times 8)$.

\subsubsection{Supersymmetry in ten dimensions}
There are various ten dimensional supersymmetric algebras, see \cite{west,so,pol} and references therein.
Recall that when $d=10$ we can define Majorana-Weyl spinors (see table \ref{tab3}). Such a spinor
has $16$ components. The structure of the supersymmetric algebra is very similar to 
the four dimensional case. We just give here, the $\{ \text{odd},\text{odd} \}$ part of the algebra.
Let $Q$ be a Majorana  spinor in ten dimensions. Then,  if we introduce a real central charge  $Z$
in addition to the Poincar\'e generators, using 
Table \ref{tab4},  and  arguments similar as in  the  previous section we get
the supersymmetric algebra in dimension ten  (since $\Gamma_\mu C_{10}^{-1}$ and $ C_{10}^{-1}$ are symmetric 
see table \ref{tab4})

\beqa
\label{10}
\left\{Q, Q^t\right\}= Z C^{-1}_{10} + P^\mu \Gamma_\mu C^{-1}_{10},
\eeqa

\noi with $C_{10}$ the $C-$matrix in dimension $10$ (see \eqref{ABC}).
From this equation, if we denote $Q_{\pm}= \frac12(1\pm\chi_{10}) Q$ the left- and right-handed components of $Q$, we obtain 

$$\left\{Q_\pm, Q_\pm^t\right\}= \frac12(1\pm \chi_{10}) P^\mu \Gamma_\mu C^{-1}_{10}, \ \ 
\left\{Q_\pm, Q_\mp^t\right\}= \frac12(1\pm \chi_{10})Z C^{-1}_{10},$$

\noi
since $\chi_{10} C_{10}^{-1}= - C_{10}^{-1} \chi_{10}$ (see eq.\eqref{chi-C})  
and  $\chi_{10} \Gamma_\mu=-\Gamma_\mu \chi_{10}$ 
(see eq.\eqref{chirality2}) with $\chi_{10}$
the chirality \eqref{chirality} matrix in ten dimensions.
In ten dimensions the fermionic part of the algebra is constituted of $N_+$  left-handed Majorana-Weyl spinors
and    $N_-$  right-handed Majorana-Weyl spinors.  
As in four dimensions the number of fermionic degrees
of freedom is at most $32$. Since a Majorana-Weyl spinor has $16$ components $N_+ + N_- \le 2$ leading to three
 different theories:

\begin{enumerate}
\item \underline{Type I supersymmetry}  We have one Majorana-Weyl supercharge, say $Q_+$ :

\beqa
\label{typeI}
\left\{Q_+, Q_+^t\right\}= \frac12(1+ \chi_{10}) P^\mu \Gamma_\mu C^{-1}_{10}.
\eeqa

\item \underline{Type IIA supersymmetry} 
  We have two Majorana-Weyl supercharges, of opposite chirality $Q_+$ and $Q_-$
(or one Majorana spinors $Q = \begin{pmatrix} Q_+ \cr Q_- \end{pmatrix}$) :

\beqa
\label{typeIIA}
\left\{Q_\pm, Q_\pm^t\right\}= \frac12(1 \pm\chi_{10}) P^\mu \Gamma_\mu C^{-1}_{10}, \ \ 
\left\{Q_+, Q_-^t\right\}= \frac12(1+ \chi_{10})Z C^{-1}_{10}.
\eeqa

\item \underline{Type IIB supersymmetry}  We have two Majorana-Weyl supercharges, of the same  chirality 
$Q_+{}_1$ and $Q_+{}_2$:

\beqa
\label{typeIIB}
\left\{Q_{+I}, Q^t_{+J}\right\}= \delta_{IJ}\frac12(1+ \chi_{10}) P^\mu \Gamma_\mu C^{-1}_{10}.
\eeqa 
\end{enumerate}

The type I (resp. IIA, IIB) theories appears naturally in string theory \cite{pol}.

\subsubsection{Supersymmetry in eleven dimensions}
Now, we give the eleven-dimensional supersymmetric algebra  (see {\it e.g.} \cite{west, pol, f}).
From table \ref{tab3}, in eleven dimensions, we take  a pseudo-Majorana spinor. Since 
such a spinor has $32$ components 
there is only one possible 
theory. Thus, in eleven dimension, the situation is  even simpler than in ten or four dimensions. Let $Q$ be
the pseudo-Majorana supercharge. If the bosonic sector is constituted only of the Poincar\'e generators, using
table \ref{tab4} the algebra is given by

\beqa
\label{11}
\left\{Q, Q^t\right\}= P^\mu \Gamma_\mu C_{11}^{-1},
\eeqa

\noi 
with $C_{11}$ the eleven-dimensional $C-$matrix (see \eqref{ABC2}). 
The interesting point of the eleven-dimensional supersymmetry 
is its simplicity. In addition, as we will see below
 lower dimensional theory can be obtained  by dimensional reduction or compactification.
Furthermore, eleven is the maximum dimension (with a signature $(1,d-1)$) where a supersymmetric theory can 
be formulated.
 Indeed, when $d=12$ the Majorana spinors have $64$ components\footnote{
In twelve dimensions with signature $(2,10)$ a Majorana-Weyl spinor has $32$
components.}.

 Finally, looking to table \ref{tab4}, one observes that some other types of central charges 
can be added to extend the 
algebra \eqref{11}. The possible central charges are real antisymmetric tensors of order two and five 
leading to the superalgebra

\beqa
\label{M}
\left\{Q, Q^t\right\}= P^\mu \Gamma_\mu C^{-1}_{11} + \frac12 Z_2^{\mu \nu} \Gamma_{\mu \nu} C^{-1}_{11} +
\frac{1}{5!} Z_5^{\mu_1 \mu_2 \mu_3 \mu_4 \mu_5} \Gamma_ {\mu_1 \mu_2 \mu_3 \mu_4 \mu_5} C_{11}^{-1}. 
\eeqa

\noi
The new central charges introduced here are rather different than the central charges considered
in four or ten dimensions. Indeed, such central charges are {\it not central}, being in 
antisymmetric tensor representations of the Lorentz algebra they are not scalar. This possibility
of tensorial central charges have been considered in \cite{ww}. In Quantum Field Theory, or in a theory
involving only elementary particles such central charges
are excluded by the  theorem of Haag-Lopuszanski-Sohnius \cite{hls} (since they are not  scalars). They become relevant
in a theory involving propagating $p-$dimensional extended objects (strings $p=1$, membranes $p=2$, {\it etc.} called 
generically branes or $p-$branes) \cite{brane}. For references on $p-$branes see {\it e.g.} \cite{west}.
\noi
The algebra \eqref{M} is the basic algebra which underlines  $M-$theory  and it involves
a membrane and an extended object of dimension five, a $5-$brane \cite{pol}.\\

\underline{Type IIA supersymmetry from eleven-dimensional supersymmetry}\\

If we denote by $M^{10}$ and $M^{11}$ the Minkowski space-time in $10$ or $11$ dimensions, we obviously
have $M^{10} \subset M^{11}$.  At the level of the algebra this reduces to
$\mathfrak{iso}(1,9) \subset \mathfrak{iso}(1,10) $.
This simple observation is  the starting point of the so-called dimensional reduction or compactification where 
a theory in $10$ dimensions is obtained from a theory in eleven  dimensions. 
Indeed, historically type IIA supersymmetry was obtained in such a process (see {\it e.g.} \cite{west}).
Starting from the eleven dimensional supersymmetry we take the eleventh dimension $x^{10}$ to be
on a circle $S^1$, and we let the radius of the  circle to be very small,  such that 
$M^{11} \to M^{10} \times S^1$. 
If we denote by $L_{MN}, P_M, 0 \le M,N\le 10 $ the Poincar\'e generators in eleven dimensions they reduce
to  $L_{\mu \nu}, P_\mu, 0\le \mu, \nu \le 9$ the generators of the $10-$dimensional Poincar\'e algebra and to
$P_{10}$ a scalar with respect to $\mathfrak{iso}(1,9)$. 
At the level of  Clifford algebras we have $\C_{1,9} \subset \C_{1,10}$. This means that  
a Majorana spinor  in eleven dimensions reduces to two Majorana-Weyl spinors $Q_+$ and $Q_-$ in ten dimensions 
because the Dirac $\Gamma-$matrices $\Gamma_{10}$ (presents in eleven dimensions)
plays the role of the chirality matrix $\chi_{10}$ in ten dimensions
(see eq \eqref{gammad}). Thus the Majorana spinor reduces to two opposite chirality Majorana-Weyl spinors
$Q_\pm = \frac12(1 \pm \chi) Q$.
Next observing that the $C-$matrices in eleven  and ten  dimensions are related as follow: 
$C_{11}= C_{10} \Gamma_{10}$, the
supersymmetric algebra in ten dimensions  obtained from the dimensional reduction of the eleven-dimensional algebra becomes

$$\left\{Q,Q^t\right\}=P^\mu \Gamma_\mu \Gamma_{10}^{-1} C_{10}^{-1} + P^{10} C_{10}^{-1}.$$

\noi which is just \eqref{10} with $Z=P^{10}$ and $\Gamma_\mu \to \Gamma_\mu \Gamma_{10}^{-1}$. 
With the two Majorana-Weyl
spinors of opposite chirality, we get the type IIA theory in ten dimensions, and $P^{10}$ becomes a
central charge.  This means that the dimensional reduction
of the eleven-dimensional supersymmetry leads to the ten dimensional supersymmetry of type IIA.\\

\underline{Four dimensional supersymmetry from eleven-dimensional supersymmetry}\\

The same principle can be applied to construct four dimensional supersymmetry
(see \cite{f} for references)
from eleven dimensional supersymmetry. But here in the compactification process we have several 
possibilities  for the compact manifold. One of the simplest compactification is to consider a $7-$sphere
such that $M^{11} \to M^{4} \times S^7$ with $M^4$ the four dimensional Minkowski space-time.

In  the reduction process from eleven  dimensions to four  dimensions we observe several things

\begin{enumerate}
\item  $\mathfrak{so}(1,10) \supset \mathfrak{so}(1,3) \times \mathfrak{so}(7)$;
\item  the Poincar\'e generators $P^M, M=1,\cdots,10$ give the Poincar\'e generators in four dimensions 
$P^\mu, \mu=0,\cdots,3$ and
$7$ scalars with respect to $\mathfrak{iso}(1,3)$ $P^4, \cdots, P^{10}$;
\item the spin representation ${\bf 32}$ of $\text{ Spin}(1,10)$ decomposes to 
${\bf 32}= ({\bf 2}_+, {\bf 8}) \oplus ({\bf 2}_-, {\bf 8})$ with ${\bf 2_\pm}$ a left/right-handed spinors of
$\mathfrak{so}(1,3)$ and  ${\bf 8}$ a Majorana spinor of $\mathfrak{so}(7)$. 
(Real spinors exist for $\text{ Spin}(1,3), \text{ Spin}(7), \text{Spin}(1,10),$ see table \ref{tab3}.)
Thus, the supercharge $Q$ gives $8$ Majorana supercharges $ Q_I$ in four dimensions;
\item the Dirac $\Gamma-$matrices decompose as follow 

$$\Gamma^M, M=0,\cdots,10, \ \to \left\{ \begin{array}{ll}
\Gamma_\mu = \Gamma^{(4)}_\mu \otimes I^{(7)}, &\mu=0,\cdots,3, \cr
\Gamma_m= i\chi_4 \otimes \Gamma^{(7)}_m,& m=1,\cdots,m \end{array}\right.,$$

\noi 
with $\Gamma^{(4)}_\mu$ the four dimensional Dirac $\Gamma-$matrices,  $\Gamma^{(7)}_m$ the matrices of the
representation of $\C_{7,0}$, $\chi_4$ the four dimensional chirality matrix and $I^{(7)}$ the identity
of $\C_{7,0}$;
\item the matrix $C_{11}$ decomposes into $C_{11}= C_{4} \otimes C_{7}$ with
 $C_4=i \Gamma_0^{(4)} \Gamma_2^{(4)}$ and $C_7= -i\Gamma_4^{(7)} \Gamma_6^{(7)} \Gamma_8^{(7)} \Gamma_{10}^{(7)}$.
\end{enumerate}

With all these observations the algebra reduces to

$$\{Q,Q^t\}= P^\mu \Gamma_\mu^{(4)} C_{4}^{-1} \otimes C_{7}^{-1} + i P^m \chi_4 C_4^{-1} \otimes \Gamma^{(7)}_m C_{7}^{-1}$$

\noi
 Next, observing that $C_7$ is symmetric (see
\eqref{ABC}) and introducing $Q_L{}_I=({\bf 2}_+, {\bf 8}),\  Q_R{}_I=  ({\bf 2}_-, {\bf 8})$ 
and $Q_I= \begin{pmatrix} Q_L{}_I \cr  Q_R{}_I \end{pmatrix}$
with $I=1,\cdots 8$ the algebra
can be rewritten (after an appropriate diagonalisation $C_7 \to I^{(7)}$)

$$\{Q_I,Q^t_J\}= \delta_{IJ}P^\mu \Gamma_\mu^{(4)} C_{4}^{-1}  + 
i P^m  \Gamma^{(7)}_m{}_I{}^K \delta_{KJ}\chi_4 C_4^{-1},$$

\noi
(with $ \Gamma^{(7)}_m{}_I{}^K$ the matrix elements of the matrices $\Gamma^{(7)}_m)$
which  is the $8-$extended supersymmetric algebra in four dimensions with central 
charge $Y_{IJ}=  P^m  \Gamma^{(7)}_m{}_I{}^K \delta_{KJ}$ (see \eqref{susy4d}).

Since the isometry group of $S^7$ is $SO(8) \supset SO(7)$, in fact in can be shown that the supercharges belong to the
vector representation  of $SO(8)$ (see \cite{f} and references therein).

\subsection{Irreducible representations of supersymmetry}
Since the Poincar\'e algebra admits a semi-direct structure $\mathfrak{iso}(1,3) = \mathfrak{so}(1,3) \ltimes 
{\cal D}_{\frac12,\frac12}$,
with ${\cal D}_{\frac12,\frac12}$ the vector representation of $\mathfrak{so}(1,3)$ ({\it i.e.} the space-time 
translations), the representations of $\mathfrak{iso}(1,3)$ are obtained by the method of induced representations  
of Wigner \cite{wi} (this also hold for any space-time dimensions).
This method consists of finding a representation of a subgroup of the Poincar\'e group and boosting it to the full group.
If we denote $P^\mu$ the momentum,  we have $P_\mu P^\mu=m^2$ for a particle of mass $m\ne 0$ and $P_\mu P^\mu=0$ for 
a massless particle\footnote{
$P_\mu P^\mu$ is a Casimir operator of the Poincar\'e algebra, its
eigenvalue is the mass.}. In practice, we go in some special frame where the momentum have a specific expression 
$P^\mu=(m,0,\cdots,0)$ in the massive case and $P^\mu=(E,0,\cdots,0,E)$ (with $E>0$)
 in the massless case. We identify the subgroup 
$H \subset ISO(1,d-1)$, called the little group,  which leaves the momentum invariant, find the representations of 
$H$ and then induce the representations
to the whole group $ISO(1,d-1)$. This method, is in fact very close to the principle of equivalence of special
relativity which states that if a result is obtained in a specific frame it can be extended to any frame of
reference.  

In this lecture, we  will only study the case of massless particles. If we denote $L_{\mu \nu}$ the generators
of the Lorentz algebra $\mathfrak{so}(1,d-1)$ the little group leaving $P^\mu=(E,0,\cdots,0,E)$ 
$(P_\mu= P^\nu \eta_{\mu \nu}= (E,0,\cdots,0,-E)$) invariant is generated by
$L_{ij}, 1 \le i < j \le d-2$ and $T_i=L_{i {0}} + L_{i \ {d-1}}, 1 \le i \le d-2$ (since $[L_{ij}, P_\mu]=0, [T_i,P_\mu]=0$).
This group is isomorphic to $E(d-2)$ the group of rotations-translations in $(d-2)$ dimensions. Since this group 
is non-compact and we are interested in finite dimensional unitary representations, we will represent the generators
$T_i$ by zero. By abuse of notations we will now call $SO(d-2)$ the little group. The frame where 
$P^\mu=(E,0\cdots,0,E)$ will be called the ``standard frame''.
This  method can be extended to study the massless representations of the supersymmetric extension of
the Poincar\'e algebra.  The little algebra now contains the bosonic generators $L_{ij}, 1 \le i < j \le d-2$ the
 central charges (if there exists) and the fermionic supercharges $Q$.

\subsubsection{Four dimensional supersymmetry}
\label{repsusy4d} 

In four dimensions, the generators of the supersymmetric algebra in the little group reduce  
$L_{12}$, $X_{IJ}, Y_{IJ}$ 
and $T_a$ for the bosonic sector (we assume here that the automorphism group of the algebra is 
$SU(N)$, (see Remark \ref{auto}) and 
$Q_I, I=1,\cdots, N$ for the fermionic sector.  
The generator $L_{12}$ is called the generator of helicity. We study here the representations of the
supersymmetric algebra when the central charge are put to zero. (In fact it can be proven that
massless particles represent trivially the central charges \cite{so}.)\\

The $\{\text{odd}, \text{odd} \}$ part of the algebra gives (in the two
components notation (see Remark \ref{2-4}))

\beqa
\left\{Q_L{}_I,Q^t_{RJ}\right\}= 2 E\delta_{IJ} ( \sigma_0 +  \sigma_3) i \sigma_2 =
4 E \delta_{IJ}\begin{pmatrix} 0 & 1 \cr 0 & 0 \end{pmatrix}.
\eeqa 

\noi
If we write $Q_L{}_I=\begin{pmatrix} Q_1{}_I \\ Q_2{}_I\end{pmatrix}$ and 
$Q_R{}_I=\begin{pmatrix} \bar Q^{\dot 1}{}_I \\ 
\bar Q^{\dot 2}_{}I\end{pmatrix}$ in the usual notations (see \cite{wb}), the 
only non-zero brackets are given by
 
$$\left\{Q_1{}_I,\bar Q^{\dot 2}{}_J\right\}= 4E \delta_{IJ}.$$

\noi Since the other brackets vanish and since we want unitary representations this means that 
$Q_2{}_I=\bar Q^{\dot 1}{}_I=0$ and the supercharges 
$a_I=\frac{Q_1{}_I}{\sqrt{4E}},\ a^\dag{}_I= \frac{\bar Q^{\dot 2}{}_I}{\sqrt{4E}}$
generate the Clifford algebra $\C_{2N,0}$ (see Remark \ref{weyl}). 
Now, the action of the bosonic part on the  supercharges 
gives $[L_{12},Q_I]= \Gamma_{12} Q_I$. Using $\Gamma_{12}=\frac12 \Gamma_1 \Gamma_2$ and \eqref{gamma4}, we get

\beqa
\label{helicity}
[L_{12}, Q_1{}_I] = -i\frac12 Q_1{}_I, \  [L_{12}, \bar Q^{\dot 2}{}_I] =i\frac12 \bar Q^{\dot 2}{}_I,
\eeqa

\noi and thus $ Q_1{}_I$ are of  helicity $\frac12$ and  
$\bar Q{}^{\dot 2}{}_I$ of helicity $-\frac12$.
 If $M_{12} \left| \lambda\right> = -i \lambda  \left| \lambda\right>$ then 
$M_{12}  Q_1{}_I\left| \lambda\right> = -i (\lambda  +\frac12) Q_1{}_I\left| \lambda\right>$
and $M_{12}Q^{\dot 2}{}_I \left| \lambda\right> = -i(\lambda- \frac12)Q{}^{\dot 2}{}_I \left| \lambda\right>$.
Finally, $Q_1{}_I$ is in the
$N-$dimensional representation of $SU(N)$  that we denote ${\bf N}$ and  $\bar Q{}^{\dot 2}{}_I$ 
is in the complex 
conjugate representation $\bar {\bf N}$. 
Thus we have $Q_1{}_I = (\frac12,{\bf N}), \ \bar Q^{\dot 2}{}_I=(-\frac12, \bar {\bf N})$ with respect
to the group $\text{Spin}(2) \times SU(N)$.

The representations of the four dimensional supersymmetric algebra are then completely  specified and is
of dimensions $2^N$, corresponding to the spinor representations of $\text{Spin}(2N)$. 
The left-handed spinors of $\text{Spin}(2N)$ will correspond {\it e.g.} to  the fermions and the
right-handed spinors to the bosons as we will see. 
We also know that $(Q_1{}_I, \bar Q^{\dot 2}{}_I)$ belongs to the vector representation of $\text{Spin}(2N)$.
But, if one wants to identify the particles content 
of the supersymmetric multiplet, it is interesting to decompose the multiplet with respect to the group
$\text{Spin}(2N) \supset \text{Spin}(2) \times SU(N)$
({\it i.e.} the little group). For the supercharge we have in this embedding 
$ {\bf 2N} = (\frac12,{\bf N}) \oplus
 (-\frac12,\bar {\bf N})$ (the supercharge are in
the vector representation ${\bf 2N}$ of $\text{Spin}(2N)$ -see Remark \ref{weyl}-). 
And using Remark \ref{weyl} it is easy to
decompose the spinor representation ${\bf 2^{N}}$ of $\text{Spin}(2N)$ into representations of 
$\text{Spin}(2) \times SU(N)$. If we 
introduce a Clifford vacuum $\Omega$ of helicity $\lambda_{\text{max}}$ ($L_{12} \Omega= -i 
\lambda_{\text {max }} \Omega$) annihilated by $a_I \ (a_I \Omega=0$)
and being in some representation of $SU(N)$ the full representation is obtained by the action of the operator of
creation $a^\dag{}_I$. For simplification we assume that $\Omega$ is in the trivial representation of $SU(N)$ 
and we obtain
the supermultiplet

\beqa
\label{multiplet-4}
\begin{array}{lll}
\ \ \ \ \ \ \ \ \text{state} & \  \text{helicity} & \text{representation of } SU(N)-\text{ dimension} \\
\\
\left|\lambda_{\text{max}} \right>& \lambda_{\text {max}} &\ \ \ \ \ \ \  [0] \ \ \ \ \ \ \ \ \ \text{ dim }= 1 \\
\bar Q^{\dot 2}{}_I \left|\lambda_{\text{max} }\right>& \lambda_{\text {max}}-\frac12 &\ \ \ \ \ \ \  [\bar 1]  
\ \ \ \ \ \ \ \ \ \text{ dim }= N \\
\bar Q^{\dot 2}{}_{I_1} \bar Q^{\dot 2}{}_{I_2}  \left|\lambda_{\text{max}}  \right>& \lambda_{\text {max}}-1 &
 \ \ \ \ \ \ \  [\bar 2]  \ \ \ \ \ \ \ \ \
  \text{ dim }= \begin{pmatrix} N \\ 2 \end{pmatrix} \\
\ \ \ \ \ \ \ \ \vdots &  \ \ \ \ \vdots & \ \ \ \ \ \ \  \ \ \  \ \ \ \ \ \ \  \vdots \\
\bar Q^{\dot 2}{}_{I_1} \bar Q^{\dot 2}{}_{I_2} \cdots \bar Q^{\dot 2}{}_{I_k} \left|\lambda_{\text{max}} \right>& 
\lambda_{\text {max}}-\frac{k}{2} & \ \ \ \ \ \ \ [\bar k]  \ \ \ \ \ \ \ \ \
 \text{ dim }= \begin{pmatrix} N \\ k \end{pmatrix}  \\
\ \ \ \ \ \ \ \ \vdots & \ \ \ \ \vdots & \ \ \ \ \ \ \  \ \ \  \ \ \ \ \ \ \  \vdots \\
\bar Q^{\dot 2}{}_{1} \bar Q^{\dot 2}{}_{2} \cdots \bar Q^{\dot 2}{}_{N} \left|\lambda_{\text{max}} \right>
& \lambda_{\text {max}}-\frac{N}{2}&\ \ \ \ \ \ \  [\bar N]  \ \ \ \ \ \ \ \
 \text{ dim }= 1
\end{array}
\eeqa

\noi with $1 \le I_1 < I_2 < \cdots < I_{N-1} \le N$, $[\bar k]$ the antisymmetric tensor of 
order $k$ of $SU(N)$ and $\begin{pmatrix} N \\k \end{pmatrix}$ its
dimension. In this decomposition we have

\beqa
\label{fermion-boson}
({\bf 2^{N-1}})_L&=& (\lambda_{\text {max }}, [0]) \oplus  (\lambda_{\text {max }} -1 , [2 ]) \oplus  
(\lambda_{\text {max }} -2 , [4 ])   \oplus \cdots  \\
({\bf 2^{N-1}})_R&=& (\lambda_{\text {max }}-\frac12 , [1]) \oplus  (\lambda_{\text {max }} -\frac32, [3 ]) \oplus \cdots  
  \nonumber 
\eeqa

\noi 
for the left/right-handed part of the spinor representation of $\text{Spin}(2N)$.
For instance if $\lambda_{\text {max}} \in \mathbb N + \frac12$ the left-handed spinors of $\text{Spin}(2N)$ are fermions 
and the
right-handed spinors are bosons. A multiplet contains bosons and fermions and a supersymmetric transformation sends a boson 
to a fermion
and {\it vice versa}. We also see that we have an equal number of bosons and fermions.

However, these irreducible representations are not enough for particle physics. 
We still have to take the CPT symmetry into account
(C means charge --or complex-- conjugation, P parity transformation and T time reversal). Under the CPT symmetry 
$Q_R \to Q^\star_R=-i \sigma_2 Q_L$ 
(or $a^\dag{}_I \to a_I$)  and $L_{12} \to -L_{12}$. A quantum field theory has to be CPT invariant.
This means if the multiplet \eqref{multiplet-4} is not invariant under this conjugation, we have to consider the CPT 
conjugate multiplet obtained by 
acting on the conjugated Clifford vacuum $\left| -\lambda_{\text{max}}\right>$ 
with the annihilation operator 
$ Q_{ 1}{}_I$ ($\bar Q^{\dot 2}{}_I\left| -\lambda_{\text{max}}\right> =0$). Let us give the 
result for certain values of $N$.

\begin{enumerate}
\item For $N=1$ the multiplet \eqref{multiplet-4} {\it are not CPT conjugate}. The particle content 
(after adding the CPT conjugate multiplet) is \\
\begin{enumerate}
\item[]  $\lambda_{\text{max}}=\frac12$
 $ \begin{array}{llll} \text{helicity}& \frac12&0&-\frac12 \cr \text{states} &1&2&\ \ \,1 \end{array}$;

(In this multiplet we have $\Omega = \left|\frac12\right>, \bar Q^{\dot 2} \Omega=\left|0\right>$,
$\Omega^{\text{CPT}} = \left|-\frac12\right>,  Q_{1} \Omega^{\text{CPT}}=\left|0\right>'$.)
\item[]   $\lambda_{\text{max}}=1$    $ \begin{array}{lllll} \text{helicity}& 1&\frac12&-\frac12& -1 \cr 
\text{states} &1&1&\ \ \, 1&\ \ \, 1 \end{array}$;
\item[]   $\lambda_{\text{max}}=2$ $ \begin{array}{lllll} 
\text{helicity}& 2&\frac32&-\frac32& -2 \cr \text{states} &1&1&\ \ \,1&\ \  \,1 \end{array}$.
\end{enumerate}
To identify the particle content of the various multiplet, we have to keep in mind
that a massless particle of spin $s$ is constituted of a state of helicity $s$ and 
a state of helicity $-s$. For instance a left-handed  massless electron is constituted of a
state of helicity $1/2$ and a state of helicity $-1/2$. The former can be interpreted as a 
left-handed electron of helicity $-\frac12$
and the latter as its corresponding anti-particle state the left-handed positron of helicity
$\frac12$.
Thus, after boosting the representations to the whole Poincar\'e group, the first  multiplet contains a 
left-handed fermion and a complex scalar field and the second 
multiplet contains a real pseudo-Majorana spinor and a real vector field. These types of multiplet are essential in the 
construction
of the so-called minimal supersymmetric standard model $-$MSSM$-$ ({\it i.e.} to construct a model describing particle
 physics and being supersymmetric).
The former multiplets, called the chiral multiplets,
are the matter multiplets (they correspond for instance to a left-handed electron and a scalar electron named selectron), 
although the
second multiplets, called the vector multiplets, are  relevant for the description of supersymmetric fundamental interactions 
(in the case of electromagnetism, it
corresponds to the photon and its fermionic supersymmetric partner the photino) \cite{wb}. 
This can be generalised for all the particles.
 This means that
in supersymmetric theory the spectrum is doubled, and to each known particle we have to add its supersymmetric partner 
\cite{MSSM}.
The last  types of multiplet (the gravity multiplet) contains the graviton
(so describes gravity) and its supersymmetric partner, a spinor-vector named the gravitino.  It is
possible to couple the gravity multiplet with the multiplet of the MSSM \cite{wb,MSSM}.
\item For $N=4$, when  $\lambda_{\text{max}}=1$ the multiplet is CPT conjugate and is not CPT conjugate for 
 $\lambda_{\text{max}}=2$:
\begin{enumerate}
\item[]  $\lambda_{\text{max}}=1$
 $ \begin{array}{llllll} \text{helicity}& 1&\frac12&0& -\frac12&-1 \cr \text{states} &1&4&6&\ \ \, 4& \ \ \, 1 \end{array}$;
\item[]
\item[]  $\lambda_{\text{max}}=2$
$ \begin{array}{llllllllll} \text{helicity}& 2&\frac32& 1&\frac12& 0&-\frac12&-1&-\frac32&-2\cr \text{states} 
&1&4&6&4& 2& \ \ \,4&\ \ \,6&\ \ \,4&\ \ \,1 \end{array}$. 

\end{enumerate} 
We observe that there is no multiplet with $\lambda_{\text{max}}=\frac12$ when $N=4$. Thus $N=4$ does not contain 
matter multiplets, but
only gauge ($\lambda_{\text{max}}=1$) or gravity ($\lambda_{\text{max}}=2$)  multiplets.
\item For $N=8$ there is only one (gravity, $\lambda_{\text{max}}=2$) multiplet which is CPT conjugate:
\begin{enumerate}
\item[]
$ \begin{array}{llllllllll} \text{helicity}& 2&\frac32& \ \, 1&\ \, \frac12& \ \, 0&-\frac12&-1&-\frac32&-2\cr 
\text{states} &1&8&28&56& 70&  \ 56& \;28&\ \ \,8&\ \ \,1 \end{array}$
\end{enumerate} 
\end{enumerate}

Several remarks are in order here

\begin{remark}
One can observe by a direct counting that in a supersymmetric multiplet there is an equal number of bosons and fermions.
This is a general result valid in any space-time dimensions \cite{wb,west}
(see also \eqref{fermion-boson}).
\end{remark}

\begin{remark}
\label{N<8}
If $N >8$ the supersymmetric multiplets contain states of helicity bigger than $2$. Since there is not consistent theory for
interacting particles of helicity bigger than $2$ in four dimensions, $N \le 8$. The complicated multiplets for $N>1$ 
can be obtained
by compactification of higher dimensional theories. For instance the four dimensional  $N=8$ extended supersymmetry 
can be obtained from the eleven-dimensional
theory or the ten dimensional type IIA or type IIB theories and the four dimensional $N=4$ extended supersymmetry 
can be obtained from the ten-dimensional type I theories
\cite{so,west,pol}.
Conversely, this compactification limits the number of supersymmetry one can take in a given dimension (see \eqref{susy-d}) 
\end{remark}

\subsubsection{Eleven dimensional supersymmetry}
In eleven dimensions, the bosonic part of the little group is $SO(9)$ and  the supersymmetric algebra  becomes

$$\left\{Q, Q^t \right\}= E(\Gamma_0 + \Gamma_{10})C^{-1}_{11}=E(1+ \Gamma_{10} \Gamma_0)\Gamma_0 C^{-1}_{11}.$$

\noi
Since the trace of $ \Gamma_{10} \Gamma_0$ is equal to zero, $(\Gamma_{10} \Gamma_0)^\dag = \Gamma_{10} \Gamma_0 $  and  
$(\Gamma_{10} \Gamma_0)^2=1$,  this means that
the matrix  $ \Gamma_{10} \Gamma_0$ has an equal number of eigenvalues $+1$ and $-1$. Thus we can chose a basis such 
that the only non zero brackets are 

\beqa
\left\{Q_a, Q_b\right\}=\delta_{ab}, a,b=1,\cdots, 16
\eeqa

\noi
and as  in four dimensions  half of the supercharges can be represented by zero. This is a general 
property of supersymmetric algebras \cite{s}. The non-zero supercharges are called the active supercharges. 
The representation of the eleven-dimensional supersymmetric algebra turns out to be the spinor representation
of $\text{Spin}(16)$ (of dimension $256$), 
but as in four dimensions to identify the precise content of the multiplet we have to 
study the embedding  $\text{Spin}(16) \supset \text{Spin}(9)$, with $\text{Spin}(9)$ the little group. 
The active supercharges are 
in the ${\bf 16}$ (spinor) representation of $ \text{Spin}(9)$ and in the  ${\bf 16}$ (vector) representation of 
$ \text{Spin}(16)$.
To identify the representation of the supersymmetric algebra we proceed in several steps. We first observe that 
in the following 
embeddings  we have the decomposition

\beqa
\begin{array}{lll}
\text{Spin}(9) &\supset\text{Spin}(8)& \supset\text{Spin}(6) \times \text{Spin}(2)  \\
\ \ \ \ {\bf 16}&= {\bf 8}_+ \oplus {\bf 8}_-& =\Big(({\bf 4}_+,\frac12) \oplus ({\bf 4}_-,-\frac12) \Big) \oplus
  \Big(({\bf 4}_+,-\frac12) \oplus ({\bf 4}_-,\frac12) \Big)
\end{array}
\eeqa

\noi
with ${\bf 8}_\pm$ a left/right handed spinor of $\text{Spin}(8)$,  $({\bf 4}_\pm$ a left/right handed spinor of 
$\text{Spin}(6)$ and
$\pm \frac12$ the eigenvalue  of  $\text{Spin}(2)$). 

Then we study explicitly the spinor representation of $\text{Spin}(8)$. 
We denote $Q_\pm = {\bf 8}_\pm$, define a Clifford vacuum  $\Omega_\pm$
for each supercharge $Q_\pm$, and decompose along the line of remark \ref{weyl}
the $Q_\pm$ into operators of creation and annihilation
(denoted $(a_+, a_+^\dag)$ and $(a_-, a_-^\dag)$)  and obtain the spinor representation  (with $Q_+$ for instance). 
The supercharges $Q_+$ belong to the vector representation ${\bf 8}_v$ of some $\text {Spin}(8)_Q$
algebra generated by $Q_+$ -see Remark \ref{weyl}-, but they also belong to the spinor representation of 
the $\text{Spin}(8)_{\text{s.t.}}$ subgroup of the little group $ \text{Spin}(9)$. Thus we firstly study
the decomposition through the embedding $\text{Spin}(8)_{\text{s.t.}} \subseteq \text {Spin}(8)_Q$
for which ${\bf 8}_+={\bf 8}_v$

\beqa
\label{spin8}
\begin{array}{lll}
\text{state}& \text{Spin}(6)& \text{Spin}(2) \\
\Omega_+& \ \ \ \ {\bf 1}&-1 \\
a^\dag_+ \Omega_+&{\ \ \ \ \bf 4}_+&-\frac12 \\
\left[a^\dag_+\right]^2  \Omega_+&\ \ \ \ {\bf 6}&\ \ \,0 \\
\left[a^\dag_+\right]^3 \Omega_+&\ \ \ \ {\bf 4}_-&\ \ \, \frac12 \\
\left[a^\dag_+\right]^4 \Omega_+&\ \ \ \ {\bf 1}&\ \ \,  1
\end{array}
\eeqa

\noi where $\left[a^\dag_+\right]^n$ means an $n-$th antisymmetric product of operators of creation.
In this decomposition, we have chosen the subgroup $\text{Spin}(6) \times \text{Spin}(2) \subset \text{Spin}(8)$
such that $a_+^\dag= ({\bf 4}_+, \frac12)$ this gives the second line in \eqref{spin8}.
Moreover,  using $ ({\bf 4}_+, \frac12) \otimes  ({\bf 4}_+, \frac12)=
 ({\bf 6}, 1) \oplus ({\bf 10}_+, 1)$ with ${\bf 6}$ the vector representation of $\text{Spin}(6)$
and ${\bf 10}_+$ the antisymmetric self-dual tensor of order three of $\text{Spin}(6)$, and oberving
that ${\bf 6}$ corresponds to the antisymmetric tensor product of spinors 
and ${\bf 10}_+$ to the symmetric tensor product of spinors, this gives the third line in \eqref{spin8}.
Similar analysis give the $\text{Spin}(6)$ content of the other line of \eqref{spin8}.
Finally  let us mention that the eigenvalues of $\text{Spin}(2)$ are normalized 
such that their sum is equal to zero. Now, it is easy to regroup the various terms and
obtain representations of $\text{Spin}(8)_{\text{s.t.}}$:

\beqa
\label{8>6+2}
\begin{array}{ll}
\text{Spin}(6) \times \text{Spin}(2) &\subset \text{Spin}(8)_{\text{s.t.}} \\
({\bf 1},-1) \oplus ({\bf 6},0) \oplus ({\bf 1},1) &= {\bf 8}_\text{v} \\
({\bf 4}_+, -\frac12) \oplus ({\bf 4}_-, \frac12)&= {\bf 8}_-
\end{array}
\eeqa

\noi
with ${\bf 8}_{\text v}$ the vector representation of $\text {Spin}(8)_{\text{s.t.}}$ and 
${\bf 8}_\pm$ the two spinor representations.
Thus in the embedding $\text{Spin}(8)_{\text{s.t.}} \subseteq \text{Spin}(8)_Q$ 
we have the following decomposition
${\bf 16} = {\bf 8}_- \oplus {\bf 8}_\text v$.
 In a similar way, acting with the supercharges $Q_-$ we have the decomposition: 
${\bf 16} = {\bf 8}_+\oplus {\bf 8}_\text v$.
To obtain now the full $\text{Spin}(16)$ representation, we just have to tensorise the representations obtained with $Q_+$ 
and $Q_-$.
First we notice 

\beqa
\label{IIA}
{\bf 8}_{\text v} \otimes {\bf 8}_{\text v} &=&{\bf 1} \ [\Phi] \oplus {\bf 35} \ [g_{i j }] \oplus {\bf 28} \ [B_{i j }] 
\nonumber \\
{\bf 8}_+ \otimes {\bf 8}_-&=& {\bf 8}_{\text v}\ [A_i] \oplus {\bf 56}_{\text v}\ [C_{ijk}] \\
{\bf 8}_{\text v} \otimes {\bf 8}_+&=& {\bf  8}_-\ [\lambda_R] \oplus {\bf 56}_-\ [\Psi_R{}^i] \nonumber \\
{\bf 8}_{\text v} \otimes {\bf 8}_-&=& {\bf  8}_+\ [\lambda_L] \oplus {\bf 56}_+\ [\Psi_L{}^i] \nonumber
\eeqa

\noi
with $1\le i,j,k \le 8$ the $\text{Spin}(8)_{\text{s.t.}}$ indices 
and $\Phi$ a scalar, $g_{i j }$ a symmetric traceless tensor, $B_{i j }$ a two-form, $A_i$ a vector, $C_{ijk}$ a three-form, 
$\lambda_L$ a left-handed spinor, $\lambda_R$ a right-handed spinor, $\Psi_L^i$ a left-handed spinor-vector and $\Psi_R^i$
a right-handed spinor-vector of $\text{Spin}(8)_{\text{s.t.}}$. 
The second decomposition comes from \eqref{tensor2} 
The third and fourth decompositions come from the triality property
of $\mathfrak{so}(8)$ $-$look to the Dynkin diagram of $\mathfrak{so}(8)$$-$.
The field in bracket $[\ \ ]$ just represents the type of field corresponding to the given  representation.
For instance $ {\bf 28} \ [B_{i j }]$ means that the $28-$dimensional representation of 
$\text{Spin}(8)_{\text{s.t.}}$  corresponds to a two-form
$B_{ij}$.
This gives the decomposition of the spinor representation of $\text{Spin}(16) \supset   
\text{Spin}(8)_{\text{s.t.}}$. To obtain now the decomposition through the 
embedding $\text{Spin}(16)\supset \text{Spin}(9)$, we just have to study
the embedding $\text{Spin}(9) \supset \text{Spin}(8)_{\text{s.t.}}$. 
If we define $I, J, K =1,\cdots,9$ the indices of $\text{Spin}(9)$
and $i,j,k=1,\cdots,8$ the indices of  $\text{Spin}(8)$, we have:

\beqa
\label{so9>so8}
{\bf 44} \ [g_{IJ}]&=& {\bf 35} \ [g_{ij}] \oplus {\bf 8}_{\text v} \ [g_{ i 9}] \oplus {\bf 1} \ [g_{ 9 9}] =
 {\bf 35} \ [g_{ij}] \oplus {\bf 8}_{\text v} \ [A_i] \oplus {\bf 1} \ [\Phi] \nonumber \\
{\bf 84} \ [C_{IJK}]&=& {\bf 56}_{\text v} \ [C_{ijk}] \oplus {\bf 28} \ [C_{ij9}] =
{\bf 56}_{\text v} \ [C_{ijk}] \oplus {\bf 28} \ [B_{ij}] \\
{\bf 128} \ [\Psi^I] &=& {\bf 56}_+\ [\Psi_L^i ] \oplus {\bf 8}_+  \ [ \Psi_L^9 ] \oplus  {\bf 56}_-\ [\Psi_R^i ] 
\oplus {\bf 8}_- \ [ \Psi_R^9 ] 
\nonumber \\
&=&
{\bf 56}_+\ [\Psi_L^i ] \oplus {\bf 8}_+  \ [\lambda_L] \oplus  {\bf 56}_-\ [ \Psi_R^i ] \oplus {\bf 8}_- \ [\lambda_R ].
\eeqa

\noi
Thus finally the representation of the eleven-dimensional supersymmetric algebra contains one spinor-vector 
(a Rarita-Schwinger field),
$\Psi^I$, a symmetric traceless second order tensor (a tensor metric) and a three-form:

\beqa
\label{11-rep}
{\bf 256}={\bf 128}_L \oplus {\bf 128}_R \ \left\{ 
\begin{array}{ll}
{\bf 128}_R=\Psi^I &\text{fermion}  \\
{\bf 128}_L=  g_{IJ}, \ \ C_{IJK}& \text{bosons}.
\end{array}\right.
\eeqa

\noi
These fields are representations of
the little group $\text {Spin}(9)$.  It becomes easy to obtain a representation of the 
Poincar\'e  algebra not limiting the indices
to their  $\text {Spin}(9)$ values but allowing  their $\text {Spin}(1,10)$ values. For instance $g_{I J}, 1 \le I,J \le 9 \to
g_{M N}, 0 \le M,N \le 10$.\\

\noi
\underline{Compactifications}\\ 

Having obtained the representation of the eleven-dimensional supersymmetric algebra we can now  obtain
the representations in smaller dimensions by compactification. Starting with the multiplet  \eqref{11-rep} 
the four dimensional representation of
the extended  $N=8$ supersymmetry can be built. In the little group  $SO(9)$ 
the indices of the fields  $I,J=1,\cdots,9 \to (i,j=1,2$ 
and $m,n=1,\cdots, 7$)
corresponding to their $\text{Spin}(2)$ and $\text{Spin}(7)$ content.
Thus we can decompose easily the fields in dimensional reduction from eleven to four dimensions. (In the simplest 
compactification, called
the trivial compactification, we simply assume that the fields do not depend on the components of the compact dimension). 
This gives

\beqa
\label{11->4}
\begin{array}{llllll}
g_{IJ} &\to& \ \ \ \ \ g_{ij} &\ \ \ \ g_{i m} &\ \ \  g_{mn} & \cr
&& 1 \text{ graviton}& 7 \text{ vectors}&28 \text{ scalars}& \cr
C_{IJK} &\to &\ \, C_{ijk}&\ \ C_{ijm}&\ \ \  C_{imn}&\ \ \ C_{mnp} \cr
&&\text{empty}&7 \text{ scalars}& 21 \text { vectors}& 35 \text{ scalars} \\
\Psi^I &\to&\ \ \ \ \ \ \Psi^i& \ \ \ \ \ \ \Psi^m \\
&&8 \text{ gravitinos} &58 \text{ spinors}
\end{array} \nonumber 
\eeqa

\noi
In this decomposition we have to pay attention because with respect to $\text{Spin}(2)$ a three-form do not exists 
and a two-form is dual to a scalar. For the fermionic part of the multiplet we have to remember that through
$\text{Spin}(9) \supset \text{Spin}(2) \times \text{Spin}(7)$, we have the decomposition of a spinor
${\bf 16}=(\frac12, {\bf 8}) \oplus (-\frac12, {\bf 8})$.  Finally, counting the number of states shows that we
obtain the $N=8$ gravity multiplet.

\begin{remark}
The so-called Yang-Mills multiplet of the ten dimensional type I supersymmetry 
can be deduced from \eqref{8>6+2}.  The compactification of the type IIA supersymmetry
from the eleven dimensional supersymmetry can be read off  \eqref{so9>so8}. 
In addition, using formul\ae \eqref{IIA} with ${\bf 8}_\pm \otimes {\bf 8}_\pm= {\bf 1} \ [\Phi] \oplus {\bf 28} 
\ [B_{ij}]\oplus {\bf 35} \ [D_{ijkl}^\pm] $ with $D_{ijkl}^\pm$ an (anti-)self-dual 
four-form of $\text{Spin}(8)$
all ten dimensional supersymmetric multiplets can be obtained:
 \begin{enumerate}
\item Yang-Mills multiplet type I: one vector $A^i$ one right-handed spinor $\lambda_R$. 
(The spinor is in the opposite chirality than the supercharge.)
\item Gravity type multiplet (the vacuum $\Omega$ is a vector of $\text{Spin}(8)$): one scalar $\Phi$,
one tensor metric $g_{ij}$ one two-form $B_{ij}$; one left-handed spinor $\lambda_L$ one left-handed
spinor-vector $\Psi_L^i$. (The spinor and spinor-vector are in the same chirality than the
supercharge).
\item type IIA one scalar $\Phi$, one tensor metric $g_{ij}$ one two-form $B_{ij}$, one vector $A_i$, one
three-form $C_{ijk}$, one left-handed and one right-handed spinor $\lambda_R, \lambda_L$
 and  one left-handed and one right-handed
spinor-vector
$\Psi_L^i, \Psi_R^i$. The fermions are of both chirality.
\item type IIB:  two scalars $\Phi, \Phi'$, one tensor metric $g_{ij}$, 
two two-forms $B_{ij}, B'_{ij}$ one anti-self-dual
four form $D_{ijkl}^-$ two left handed spinors $\lambda_L,\lambda'_L$
 two left-handed
spinor-vector $\Psi_L^i, \Psi'^i_L$. 
\end{enumerate}
Finally let us mention that type IIA or type IIB supersymmetry
give by compactification in four dimensions $N=8$ extended-supersymmetry.
\end{remark}
\section{Conclusion}
In this lecture we have shown, that  the basic tools to construct  
supersymmetric extensions of the Poincar\'e algebra
are  Clifford algebras. Special attention have been given to the four, ten and eleven dimensional spaces-times.
Studying representations of supersymmetric algebras shows that a supermultiplet contains an 
equal number of bosonic and fermionic degrees of freedom and that supersymmetry is a symmetry which
mixes non-trivially bosons and fermions.  The next step is to apply supersymmetric theories
in Quantum Field Theory or particle physics. For that purpose we need first
to  calculate   transformation of the fields under supersymmetry  and 
then to build invariant Lagrangians (the concept of superspace is central
for this construction).  For instance all the technics of
supersymmetry have  been  applied in four dimensions to construct a supersymmetric version
of the standard model \cite{wb, MSSM}.  There are some strong arguments in favor of
such a theory, even if there is no experimental evidence of supersymmetry.
Supersymmetry or more precisely its local version
contains gravity and as such is named supergravity. Ten-dimensional supergravities appear as some low 
energy limits of string theories  and present some interesting duality properties \cite{pol,west}.
Finally, with the brane revolution a lot of hope have been put to the so-called M-theory whose
low energy limits contains the eleven-dimensional supergravity and the various strings theory \cite{pol,west}.

\paragraph*{Acknowledgements}
The Organizing Committee, and in particular P. Angles, are kindly acknowledged for the friendly and
studious atmosphere during the conference.

\end{document}